\documentclass[prb, notitlepage]{revtex4-1}
\usepackage[utf8]{inputenc}
\usepackage{physics}
\usepackage{amsmath}
\usepackage{amssymb}
\usepackage{graphicx}
\usepackage{fancyhdr}
\usepackage{pdfpages}
\usepackage{float}
\usepackage{xcolor}
\colorlet{BLACK}{black}
\usepackage{natbib}
\usepackage[colorlinks=true,citecolor=blue,linkcolor=blue,urlcolor=blue,linktocpage]{hyperref}
\usepackage{mathdots}
\usepackage[intlimits]{mathtools}
\usepackage{ulem}
\usepackage{geometry}
\usepackage{minitoc}
\usepackage{setspace}

\makeatletter
\AtBeginDocument{\let\LS@rot\@undefined}
\makeatother
\newcommand{\black}{\textcolor{black}}

\newcommand{\appropto}{\mathrel{\vcenter{
  \offinterlineskip\halign{\hfil$##$\cr
    \propto\cr\noalign{\kern2pt}\sim\cr\noalign{\kern-2pt}}}}}
\newcommand{\biggg}{\bBigg@{4}}

\newcommand{\mbf}[1]{\mathbf{#1}}

\usepackage{ulem}   
        
\usepackage{etoolbox} 
\usepackage{lipsum} 
\usepackage[capitalize]{cleveref}

\makeatletter
\appto{\appendix}{%
  \@ifstar{\def\theequation@prefix{A.}}%
          {}%
}
\numberwithin{equation}{section}

\begin{document}
\raggedbottom
\setlength{\parskip}{0pt}
\setlength\parindent{0pt}
\title{\textbf{Landau Levels, Edge States, and Gauge Choice in 2D Quantum Dots}}
\author{Asadullah Bhuiyan and Frank Marsiglio}
\affiliation{Department of Physics, University of Alberta, Edmonton, Alberta, Canada, T6G~2E1}

\begin{abstract}
We examine the behaviour of a charged particle in a two dimensional quantum dot, in the presence of a magnetic field. Emphasis is placed on the high magnetic field regime. Confinement in a dot geometry provides a more realistic system where edge effects arise naturally. It also
serves to remove
the otherwise infinite degeneracy; nonetheless, as described in this paper, additional ingredients are required to produce sensible results. We treat both circular and square geometries, and in the latter
we explicitly demonstrate the gauge invariance of the energy levels and wave function amplitudes. The characteristics of bulk states closely resemble those of 
free space states. For edge states,
with sufficiently high quantum numbers we achieve significant differences in the square and circular geometries. Both circular and square geometries are shown to exhibit level crossing phenomena, similar to parabolic dots. Confinement effects on the probability current are also analyzed; it is the edge states that contribute non-zero current to the system. Results are achieved using straightforward matrix mechanics, in
a manner that is accessible to novices in the field. On a more pedagogical note, we also provide a thorough review of the theory of single electron Landau levels in free space, and illustrate how the introduction of surfaces naturally leads to a more physically transparent description of a charged particle in a magnetic field.
\end{abstract}

\date{\today} 
\maketitle
\begin{spacing}{1}
\tableofcontents
\end{spacing}

\section{\black{Introduction}}

The motion of a charged particle in the presence of a magnetic field is a subject of considerable interest in many areas of physics. In particular, in condensed matter the presence of the magnetic field alters the particle's behaviour in a profound way. Ignoring the spin degree of freedom, it is known from classical physics that a vector potential is required to include the effect of a magnetic field, and more than one choice of vector potential is possible. Each of these choices constitutes a particular gauge. For a uniform magnetic field, say in the $z$ direction,  common choices in Cartesian coordinates are the Landau Gauge, $\vec{\bf A}_1= xB \hat{\bf y}$ (or $\vec{\bf A}_2 = -yB \hat{\bf x}$), or the so-called Symmetric gauge, $\vec{\bf A}_S = -yB/2 \hat{\bf x} + xB/2 \hat{\bf y}$; all of these choices yield the same constant magnetic field $\vec{\bf B} \equiv \vec{\bf \nabla} \times \vec{\bf A} = B {\hat{\bf z}}$. 

Naturally all physical quantities obtained in a calculation should be independent of the gauge choice. Typically a gauge choice is made to take advantage of some symmetry in the problem, so that one can proceed analytically. Just as often the physical system is altered to take advantage of the symmetry afforded by the gauge choice. A good example is the 
seminal paper by Laughlin,~\cite{laughlin1981quantized} where he wanted to address the quantization of the Hall conductivity in a two-dimensional metal. 
To this end he chose a
ribbon bent into a loop (i.e. open boundary conditions in one direction and periodic boundary conditions in the other). 
The magnetic field is perpendicular to the plane of the metal, which implies the physically questionable notion of a field either emanating from or converging to an axis at the centre of the loop. 
That this configuration did not coincide precisely with the experiment was immaterial --- the concept of quantization was properly established and what mattered was that the Landau gauge allowed Laughlin to proceed analytically with the argument. 
Similarly, Halperin~\cite{halperin1982quantized} followed up with a thin film with annular geometry, and adopted the gauge choice $\vec{\bf A} = [B_0r/2 + \Phi/(2 \pi r)] \hat{\bf \theta}$ to produce a uniform field $B_0$ in the $z$-direction along with a central flux of magnitude $\Phi$.

Other ``physical'' geometries, suitable to so-called ``quantum dots,'' were adopted thereafter --- see for example, Refs. [\onlinecite{robnik86}] and [\onlinecite{lent91}]. In this paper we will review and expand upon some of these calculations, representing charged particles in a confined geometry subjected to a constant magnetic field.
A very good reference for the quantum
dot work is Chakraborty's book, {\it Quantum Dots},\cite{chakraborty99} which also contains a number of very useful reprinted articles. As mentioned above, we will focus on the orbital motion of the charged particle (hereafter these will be electrons) and neglect the spin degree of freedom. Spin will cause a breaking of degeneracy in the presence of a magnetic field, owing to the direct coupling with field through the so-called Zeeman term. 

We begin in the next section with a textbook review of the particle in free space; this is where the idea of Landau levels first arose. Most of the material in this section is relegated to an Appendix, where we highlight both the Landau and the Symmetric gauges, and seemingly derive very different results for the wave functions. We further note that confinement within a parabolic trap can also be treated analytically, following for example, Rontani.\cite{rontani99} Next we will review and expand upon Lent's treatment of the circular quantum dot,\cite{lent91} and illustrate how this problem can be tackled with undergraduate tools. Finally, we will show numerical results for the square quantum dot in both aforementioned gauges. This will hopefully make clear the role of degeneracy, and we explicitly illustrate the equivalence of the results in the two gauge formulations by removing the degeneracy. We will present the results of calculations for simple properties only, such as probability density and current density.
Having obtained the exact energy spectrum along with the eigenstates of course allows one to compute any desired single particle property.

\section{\black{Review of Landau Levels in Free Space}}

In the presence of a magnetic field $\mathbf{B}$, the canonical momentum $\mathbf{p}$ is shifted by the magnetic vector potential $\mathbf{A}$ to give us a Hamiltonian of the form:
\begin{equation}
    H=\frac{\boldsymbol{\pi}^2}{2m}=\frac{(\mathbf{p}-q\mathbf{A})^2}{2m},
    \label{ham_original}
\end{equation}
where $q$ is the charge and $m$ is the mass of the particle of interest, and $\boldsymbol{\pi}=\mathbf{p}-q\mathbf{A}$ is the \textit{kinetic} momentum operator. Hereafter we adopt $q = -e$ for the electron, where $e>0$ is the magnitude of the charge of the electron. It is important to note that in the presence of a magnetic field, $\boldsymbol{\pi}$ represents the true momentum of the particle rather than $\mathbf{p}$. In this problem, we consider an electron free to move in a 2-dimensional system in a uniform magnetic field, 
$\mathbf{B}=B\boldsymbol{\hat{z}}$. For such a magnetic field, there are three common gauge choices, as already mentioned in the introduction. The two Landau gauges are handled very similarly mathematically, and so we will focus only on $\mathbf{A}_1$. However, $\mathbf{A}_1$ and $\mathbf{A}_S$ are handled very differently mathematically and we now discuss each in turn. A number of resources are available that provide considerable detail for this problem. These include textbooks like Refs. [\onlinecite{jain2007composite}], [\onlinecite{cohen-tannoudji77}], [\onlinecite{ballentine15}] and [\onlinecite{landau77}] and online lecture notes by Tong\cite{qheTong} and Murayama.\cite{Murayama} Systematic derivations of the eigenvalues and eigenstates for each gauge are provided in the Appendix.

A succinct summary of the Appendix is as follows. The eigenvalues are infinitely degenerate, and we obtain the same
standard expression in either gauge,
\begin{equation}
E_{n_L} = \hbar \omega_c\left(n_L + \frac{1}{2}\right), \ n_L=0,1,2,...
\label{eigen}
\end{equation}
where $\omega_c \equiv eB/m$ is the classical cyclotron frequency and $n_L$ is the Landau level quantum number. On the other hand, the wave functions in the two gauges are very different looking,
Eq.~(\ref{psi_rad}) for the Symmetric gauge, and Eq.~(\ref{psi_landau}) for the Landau gauge. In both cases there are ``hidden'' quantum numbers, $\ell$ for the Symmetric gauge and $k_y$ for the Landau gauge.

The derivations in the Appendix highlight some troublesome issues.
The infinite degeneracy is difficult, but even more so is the dependency of the wave function on gauge choice. We have encountered some gauge invariant
properties in the free space system, for example, the spectrum of energies. The energies are left invariant because a change in gauge choice $\bf A\rightarrow \bf A'$ is equivalent to performing a unitary operation on the Hamiltonian:
\begin{equation}
    \mbf{A}\rightarrow \mbf{A}'= \mbf{A} + \mbf{\nabla}\lambda  \Longrightarrow H\left[\mbf{r},\mbf{p}-q\mbf{A}'\right] = e^{iq\lambda/\hbar}H\left[\mbf{r},\mbf{p}-q \mbf{A}\right]e^{-iq\lambda/\hbar}
    \label{Ham_transform}
\end{equation}
where $\lambda$ is an arbitrary real function of position and time, and $q$ is the charge of the particle.
The wave function, on the other hand, is {\it not} gauge invariant, and for a change in gauge, $\bf A
\rightarrow \bf A'$, there is a corresponding expected change in wave function, \cite{griffiths18}
\begin{equation}
 \mbf{A}'= \mbf{A} + \mbf{\nabla}\lambda  \Longrightarrow \Psi^\prime = e^{-iq\lambda/\hbar} \Psi.
\label{psi_transform}
\end{equation}
But this means that the probability density, $|\Psi|^2$ is expected to be gauge invariant. Yet, even within the symmetric gauge, Eq.~(\ref{gauge_as}), different choices of 
origin $(x_0,y_0)$ seemingly result in wave functions that differ by far more
than a phase (and therefore probability densities that {\it do differ} from one another). 
The reason for this problem is that a degeneracy exists in the solution. An explicit calculation
of the wave function transformation corresponding to that in Eq.~~(\ref{psi_transform}) between the Landau gauge and the Symmetric gauge has been given recently in Ref.~[\onlinecite{wakamatsu18}], and this problem was first highlighted in Ref.~[\onlinecite{swenson89}], where the author presented a more general gauge condition to replace
the simple wave function relation in Eq.~~(\ref{psi_transform}) when degeneracy is present. We will return to this issue later.
The infinite degeneracy in both cases allows the wave functions in the two gauges to differ by more than the simple ``textbook'' phase, as discussed by Swenson \cite{swenson89} and Wakamatsu et al.\cite{wakamatsu18}

\begin{figure}[H]
    \centering
    \includegraphics[scale=0.5]{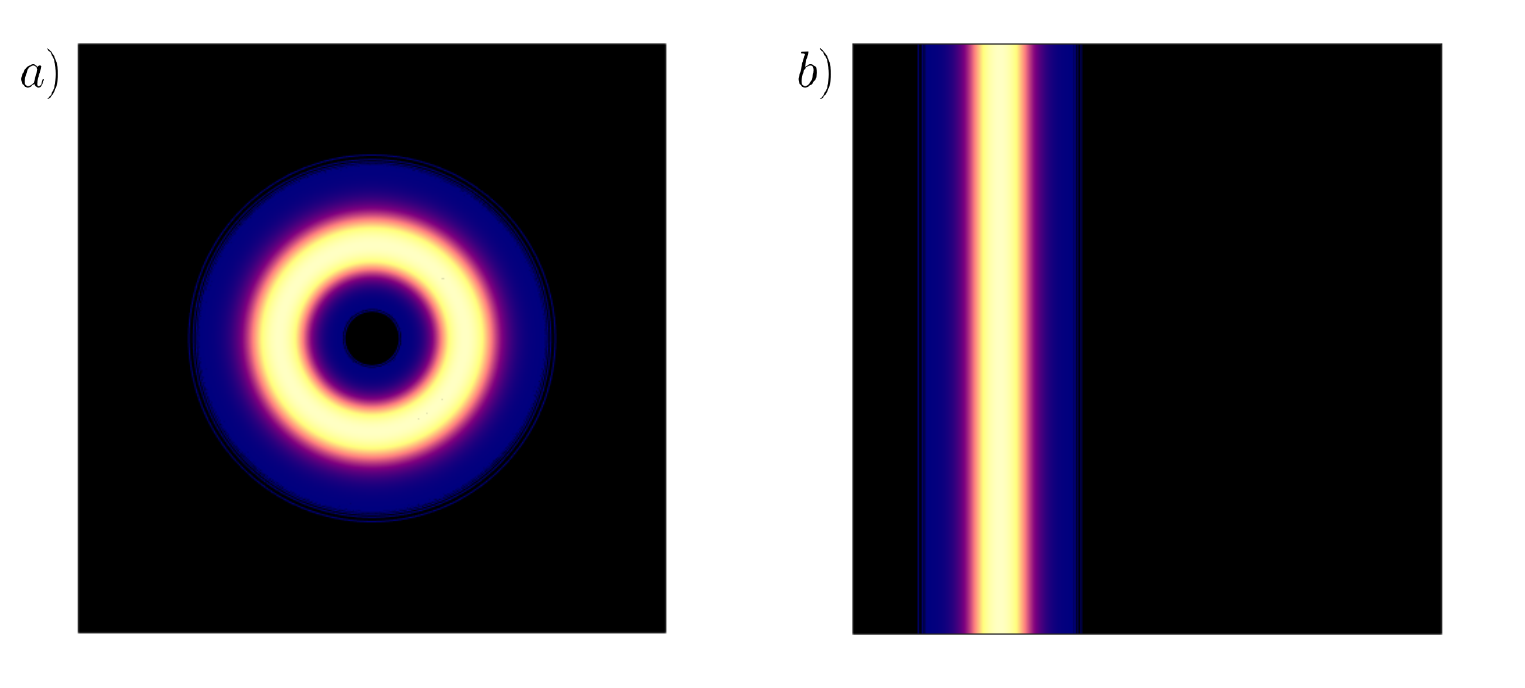}
    \caption{Contour plots of the free space probability density vs. $x$ and $y$, in length units of $l_B$, for (a) the Symmetric gauge centred about the origin, with
    $\ell=5$ and (b) the Landau gauge, with $k_yl_B=5$. The probability densities in each gauge have strikingly different spatial structure. It is clear that these eigenstates are not related by just a simple phase factor.}
    \label{fig: fs symm vs landau}
\end{figure}

To summarize,
the two gauges discussed in the Appendix both represent the same uniform magnetic field $\mathbf{B}=B\boldsymbol{\hat{z}}$. They result in the same energy spectrum, and yet they produce two very different-looking eigenstates (see Fig.~\ref{fig: fs symm vs landau}). There are gauge-invariant similarities as well; both gauges produce infinitely degenerate Landau levels in free space and both gauges have degeneracies that are expected to be approximately $G \equiv BA/\Phi_0$ in a finite space with area $A$. Eq.~(\ref{psi_transform}) suggests that the two eigenstates should differ by a simple phase factor. That does not happen here, because of the degeneracy associated with the eigenstates. The dilemma is solved by taking an appropriate linear combination of the degenerate eigenstates (which is still an eigenstate), and these can be chosen to result in a state that differs from one in the other gauge by a simple phase factor, as illustrated in Refs.~[\onlinecite{swenson89}] and [\onlinecite{wakamatsu18}].
We will encounter this dilemma once more in a situation where the degeneracy has (nominally) been lifted, and outline a resolution by means different than that used in
Refs.~[\onlinecite{swenson89}] and [\onlinecite{wakamatsu18}].

\section{\black{Landau Levels in a Circular Quantum Dot}}
\label{circ}
We will revisit the issue of degeneracy in different gauges in Section \ref{square} when we discuss an electron confined to a square dot. First, however, partly to review results due to Lent,\cite{lent91} and partly to introduce some technical aspects and more general results to be used later, we present a discussion of results for the circular quantum dot. We will use dimensionless units wherever possible. However, it is important to convert our results to physical values. As we have already seen, the flux quantum plays an important role in these problems; its value in SI units is $h/e \approx 4140$ T\ (nm)$^2$. We will use a dimensionless measure of flux, $G = BA/(h/e)$, where the scalar $A$ is the area of the sample through which a (constant) magnetic field penetrates. For circular geometry, we use $\gamma \equiv B\pi a^2/(h/e)$, where $a$ is the radius. So, for example, applying a $10$ T magnetic field
through a sample with radius 100 nm gives $\gamma \approx 76$.

\subsection{\black{Matrix Element Calculation}}

To study the effects of hard-wall confinement on Landau levels in a circular quantum dot, we use the rotationally invariant Symmetric gauge, a natural choice for such a confinement geometry. Here, our matrix mechanics scheme is outlined before moving on to numerical results.
Matrix mechanics techniques have been utilized for problems like this before, albeit for a study of disordered systems, with a different gauge and confining
potential.\cite{ohtsuki1988numerical} Otherwise, they have been used more recently in more pedagogical contexts in a variety of contexts, e.g. in Refs.~[\onlinecite{marsiglio09}, \onlinecite{jugdutt13}, \onlinecite{pavelich16}]\\[5pt]
The Hamiltonian for this system is given by:
\begin{equation}
        H = \frac{\mathbf{p}^2}{2m} + \frac{\omega_c}{2}L_z+\frac{1}{8}m\omega_c^2r^2 + V(r).
        \label{ham_circ}
\end{equation}
where our confining potential $V(r)$ is defined as:
\begin{equation}
    V(r) = \begin{cases}
            0, \ \text{if} \ r\leq a\\
            \infty, \ \text{else,}
            \end{cases}
\end{equation}\\[5pt]
where $a$ is the radius of the circular dot.
To solve the Schr\"odinger equation with this Hamiltonian, we follow the usual procedure of expanding in a basis (see, for example, 
Refs.~[\onlinecite{marsiglio09}, \onlinecite{jugdutt13}]). Our basis states of choice are the eigenstates of the infinite circular well, a natural basis for our confining geometry. These eigenstates are well known:\\[3pt]
\begin{equation}
    \braket{r,\phi}{n,\ell}=\frac{e^{i\ell\phi}}{\sqrt{2\pi}}\frac{\sqrt{2}}{a}\frac{J_\ell\left(\beta_{n,\ell}{r/a}\right)}{J_{\ell+1}(\beta_{n,\ell})}\label{circ_basis}
\end{equation}\\[3pt]
where $n=1,2,3,...$, $\ell = 0,\pm 1, \pm 2,...$, and $\beta_{n,\ell}$ are the $n$th zeros, in ascending order, of the $\ell$th order Bessel function. The corresponding eigenenergies are given by:
\begin{equation}
    E_{n,\ell} = \frac{\hbar^2\beta_{n,\ell}^2}{2ma^2}. \label{Bessel_Energy}
\end{equation}
Carrying out the standard procedure for numerical matrix mechanics, we expand the eigenstate $\ket{\Psi}$ of the Hamiltonian (Eq.~ \ref{ham_circ}) in the circular well basis: 
\begin{equation}
  \ket{\Psi}=\sum_{n^\prime=1}^{\infty}\sum_{\ell^\prime= -\infty}^{\infty}c_{n^\prime,\ell^\prime} \ket{n^\prime,\ell^\prime}
\end{equation}
where $\ket{n',\ell'}$ is the ket corresponding to the circular well basis state, Eq.~~(\ref{circ_basis}). Acting on this state with the Hamiltonian, Eq.~~(\ref{ham_circ}), and then taking the inner product of both sides with some arbitrary bra basis state $\bra{n,\ell}$ results in the following matrix equation:
\begin{equation}
    \sum_{n^\prime=1}^{\infty}\sum_{\ell^\prime=-\infty}^{\infty} H_{n \ell, n^\prime \ell^\prime} c_{n^\prime,\ell^\prime} =E_{n,\ell}c_{n,\ell}
    \label{matrix1}
\end{equation}
where 
\begin{equation}
H_{n \ell, n^\prime,\ell^\prime}=\bra{n,\ell}H\ket{n^\prime,\ell^\prime}=\delta_{n,n^\prime}\delta_{\ell,\ell^\prime}E^{(0)}_{n,l}+\frac{1}{2}\omega_c\bra{n,\ell}L_z\ket{n^\prime,\ell^\prime}+\frac{1}{8}m\omega_c^2\bra{n,\ell}r^2\ket{n^\prime,\ell^\prime}
\label{hmatrix}
\end{equation}
are the matrix elements of the Hamiltonian given by Eq.~(\ref{ham_circ}).
Here, $E_{n,l}^{(0)}$ is simply the infinite circular well eigenenergy Eq.~(\ref{Bessel_Energy}) and $\delta_{ij}$ is the standard Kronecker delta function. It should now be apparent that the entire matrix is diagonal
in $\ell$, i.e. $ H_{n \ell,n^\prime \ell^\prime} = \delta_{\ell, \ell^\prime} H_{n,n^\prime}$, and the eigenvalue problem {\it for a given $\ell$ only} needs be solved. While a diagonalization is now required for each $\ell$ separately, this greatly reduces the computational cost.
Evaluating the last two inner products of Eq.~~(\ref{hmatrix}), we obtain all the matrix elements for the Hamiltonian, 
\begin{equation}
    H_{n,n^\prime}=\delta_{n,n^\prime}\left[\frac{\hbar^2\beta_{n,\ell}^2}{2ma^2}+\frac{\hbar\omega_c\ell}{ {2}}\right]+\frac{1}{8}m\omega_c^2a^2\rho^2_{n,n^\prime}.
\label{hmatrix3}
\end{equation}
Note that we have suppressed the dependency on $\ell$ in the matrix labels, but it is carried as a parameter in all of these matrices.
Using energy units of  $\hbar^2/(2ma^2)$, we can write this in dimensionless form
\begin{equation}
    \frac{H_{n,n^\prime}}{\hbar^2/(2ma^2)} \equiv h_{n,n^\prime} = \delta_{n,n^\prime}\left[\beta_{n,\ell}^2+2{\gamma} \ell \right]+
    {\gamma^2}\rho^2_{n,n^\prime}
    \label{hmatrix4}
\end{equation}
with
\begin{equation}
    \rho^2_{n,n^\prime}=\frac{\bra{n}r^2\ket{n^\prime}}{a^2}=\frac{2\int_{0}^1J_\ell(\beta_{n^\prime,\ell}\rho)\rho^2J_\ell(\beta_{n,\ell}\rho)\rho d\rho}{J_{\ell+1}(\beta_{n^\prime,\ell})J_{\ell+1}(\beta_{n,\ell})}.
    \label{rho_defn}
\end{equation}\\
The integral in Eq.~~(\ref{rho_defn}) is non-elementary but can be evaluated analytically; we get 
\begin{equation}
 \rho^2_{n,n^\prime} = \delta_{n,n^\prime} \Bigl[ \frac{\beta^2_{n,\ell} + 2\ell^2 -2} {3 \beta^2_{n,\ell}} \Bigr] + (1 - \delta_{n,n^\prime}) \Bigl[\frac{8\beta_{n^\prime,\ell}\beta_{n,\ell}}{(\beta_{n^\prime,\ell}^2-\beta_{n,\ell}^2)^2} \frac{J_{\ell-1}(\beta_{n^\prime,\ell})J_{\ell-1}(\beta_{n,\ell})}{J_{\ell+1}(\beta_{n^\prime,\ell})J_{\ell+1}(\beta_{n,\ell})} \Bigr],
\label{bess_int1}
\end{equation}
where $\gamma$ was earlier defined as
\begin{equation}
    \gamma = \frac{B\pi a^2}{h/e}=\frac{\Phi}{\Phi_0}.
    \label{gamma_defn}
\end{equation}
We expect $\gamma$ to give us a rough measure of the degeneracy in the confined system as it is defined as the ratio of magnetic flux to flux quanta. Recall that a constant $G=\frac{\Phi}{\Phi_0}$ (Eqs.~\ref{degeneracy}, \ref{degeneracy_landau}) was argued to be a measure of degeneracy in confined system. Here, $\gamma$ plays this role.

We now have all the matrix elements analytically, to insert into Eq.~(\ref{matrix1}) for diagonalization.
The big advantage of having adopted the symmetric gauge with this geometry is that the resulting Hamiltonian is diagonal in one of the quantum numbers, $\ell$. This represents
a tremendous computational gain. We mention it here because we will {\it not} have this possibility when we study the square geometry in the next section. There we will have to diagonalize $N^2 \times N^2$ matrices, where $N$ is a large number (here typically something like $50$ or $200$. Also, for circular geometry we can monitor the quantum number $\ell$ precisely, which means we can track the radial quantum number $n_r$ as well. In the Symmetric gauge the Landau level quantum number $n_L$ is given simply by $n_L = n_r + (\ell + |\ell|)/2$.
We will not have this possibility with the square geometry.\\
To be clear, in practice in this section we diagonalize $N \times N$ matrices, and only show results that have converged as $N$ increases (recall that $N$ should be infinite, but is chosen to be finite and represents the number of radial basis states (i.e. number of $n$) for a given $\ell$ in Eq.~(\ref{circ_basis}) to be used in the matrix diagonalization).
We will do this for a number of different $\ell$'s (typically $N$ of these), again so that all presented properties are converged as a function of basis size.

\subsection{\black{Eigenenergy Spectra}}

\begin{figure}[H]
\begin{center}
\includegraphics[scale=0.3]{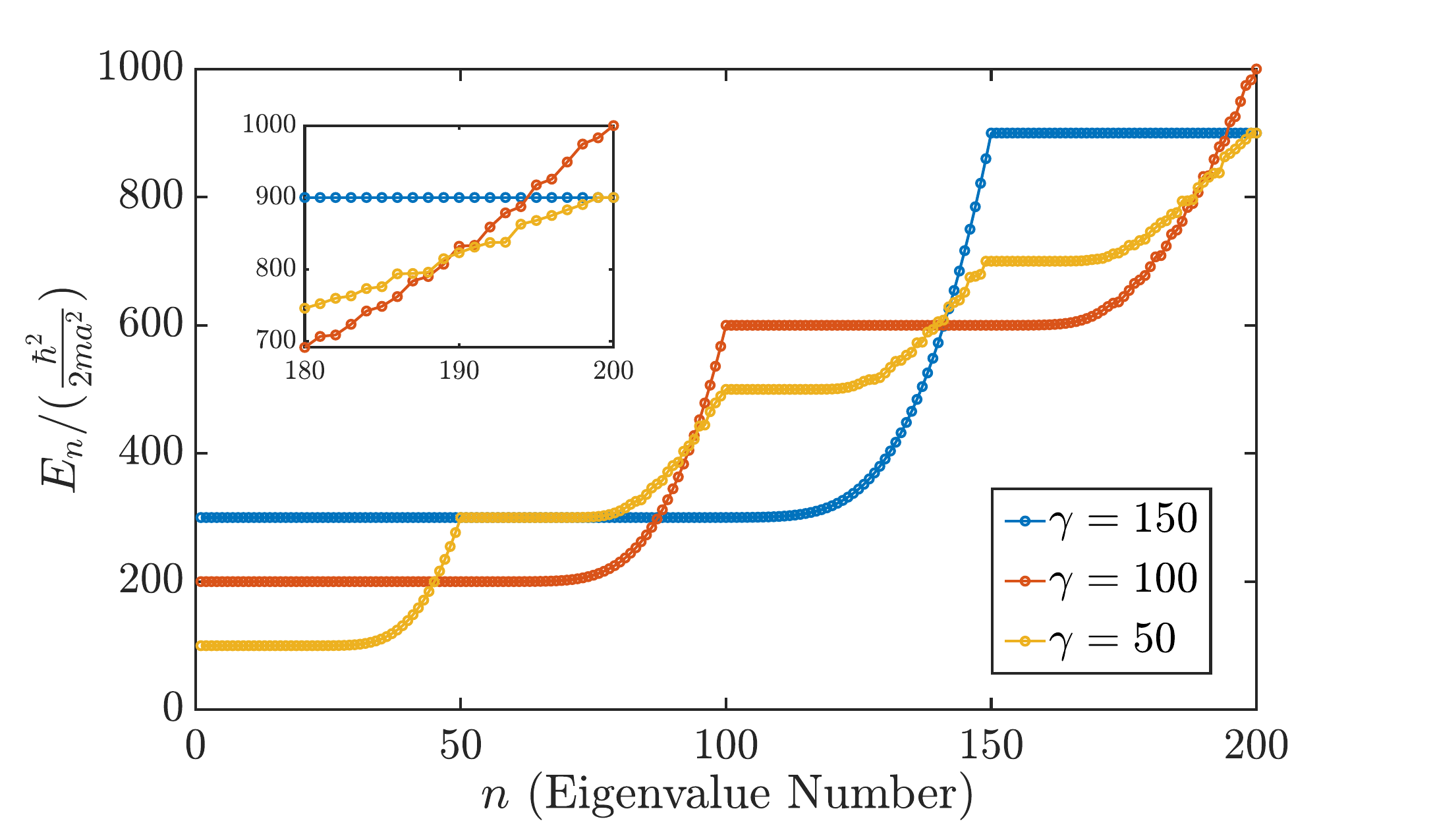}
\end{center}
\caption{Dimensionless eigenvalues ($E_n/(\hbar^2/(2ma^2))$) as a function of eigenvalue number $n$ for various values of $\gamma$. Here, $n$ is 
used as a label for the horizontal axis to indicate the energies are plotted in ascending order and does \textbf{not} refer to any previously defined quantum numbers also labelled with $n$. The truncation sizes of the matrices produced are given by $N = 200$ in 
all cases.  Note that these plots were produced as a result of generating and diagonalizing matrices for $\ell\in [-200,200]$ followed by aggregating the resulting 
eigenenergies acquired from all matrices and organizing them in ascending order. This results in over 80000 eigenvalues, and the lowest 200 of these for each value of $\gamma$, are shown here.
Note that the degeneracy number increases with the value of $\gamma$ according to 
our expectations, and in fact there are approximately (precisely for the LLL energies) $\gamma$ eigenvalues occurring before the next plateau begins. These plateaus occur at 
dimensionless energies given by those expected from the free space Landau levels: $E_{n}/(\hbar^2/(2ma^2)) = 2\gamma(2n_L+1)$, for $n_L = 0, 1, 2...$. Shown in the inset are energies occurring after the first, second and fourth plateau for $\gamma = 50, 100$ and $150$, respectively. Notice the disorderly behaviour of the energies for $\gamma = 50,100$, whereas energies between the first and second plateaus for these curves are ordered quite smoothly.}
\label{circular_eigen}
\end{figure}

Having carried out this programme, we can now order the eigenvalues in increasing value. We do this to mimic the result we would have achieved had we not recognized that
the matrix is diagonal in $\ell$ (as will be the case with the square geometry). The eigenvalue results for a few values of $\gamma$ are shown in Fig.~\ref{circular_eigen}, and show in all cases
plateaus (i.e. degenerate levels) separated by regions in which the energies increase to the next plateau. The number of points from the start of one 
plateau to the start of the next plateau is essentially $\gamma$.
The so-called ``plateaus'' are actually not degenerate to infinite precision, but they are degenerate to the precision of our computer (16-digit accuracy). These plateaus appear to be the Landau levels, originally infinitely degenerate, but now with a degeneracy of order $\gamma$, and in practice less. The increase in eigenvalues towards the end of the plateau regions is indicative that these eigenfunctions can feel the edge of the confining potential, and for this reason they are usually referred to as ``edge states'' because, as we shall see, their probability density is concentrated there. One might ask, where have the rest of the (originally) infinite states in a given Landau level gone? Fig.~\ref{circular_eigen} gives the impression that there were $\gamma$ such states, and about three quarters of them have remained degenerate, while the remaining one quarter have had their energies elevated due to the edge.

\begin{figure}[H]
\begin{center}
\includegraphics[scale=0.45]{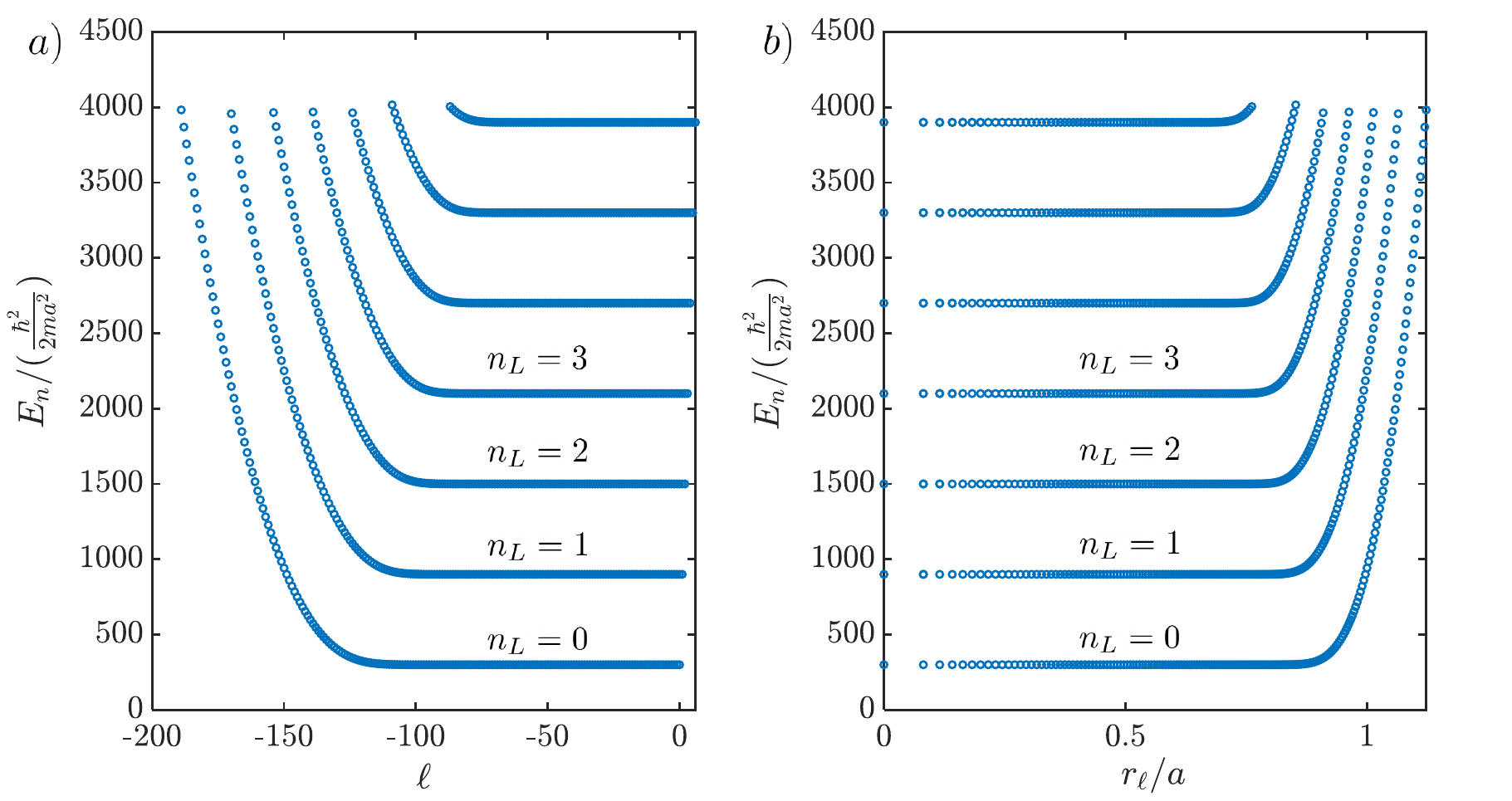}
\end{center}
\caption{Lowest 1000 Eigenenergies plotted vs. (a) $\ell$ and (b) $r_\ell/a=\sqrt{|\ell|/\gamma}$, for $\gamma=150$ and matrix truncation size, $\text{N}=200$. In both plots, each smooth energy band corresponds to a distinct Landau quantum number $n_L=n_r+(|\ell|+\ell)/2$; we labelled the first 4 to emphasize this.
$\ell$ is the 
angular momentum quantum number and $r_\ell$ is the centre of the probability density of the corresponding eigenstate as long as the eigenstate is not too 
close to the walls. This means that $r_\ell$ gives us a good idea about where the eigenstates corresponding to the eigenergies are located in real space, as long
as $r_\ell/a < 0.9$.
Energies with $r_\ell/a>1$ should be interpreted as corresponding to eigenstates piling up near the boundary. This plot (as opposed
to Fig.~\ref{circular_eigen}), makes clear that the original Landau level with infinite degeneracy ({\it all} negative values of $\ell$) has the very large $|\ell|$ states pile
up near the boundary, with increasingly higher energies, thus breaking the original free Landau (infinite) degeneracy. As is clear from the figure, though, a quasi-degeneracy
of order $\gamma$ (proportional to the applied magnetic field) remains. Notice that in a), there is an additional positive $\ell$ state as we jump from one Landau level to the next. This is in agreement with the expression derived for the symmetric gauge energies in free space. }
\label{circular_eigen_ell}
\end{figure}

A clue that this interpretation is incorrect arises from something barely discernible on this plot but clearly visible in the inset, which is that the increase in eigenvalues at the ends of the plateaus become increasingly less smooth for successive plateaus.
The reason for this is made clear in Fig.~(\ref{circular_eigen_ell}a), where now the energy levels are plotted as a function of $\ell$. Again we emphasize that we are fortunate 
that this is possible here, because different matrices were diagonalized for individual values of $\ell$. 
In Fig.~(\ref{circular_eigen_ell}b) we show the same energy 
levels vs. $r_\ell/a \equiv \sqrt{|\ell|/\gamma}$,
to show how the maximum in the wave function moves towards the edge as $\ell$ increases (with the cautionary proviso that $r_\ell$ represents the maximum only for cases well
away from the wall -- which is why $r_\ell/a$ ends up exceeding unity in this plot --- but the qualitative trend is correctly portrayed).
Figure~(\ref{circular_eigen_ell}) illustrates that the eigenenergies are smoothly increasing when plotted vs. $\ell$. The raggedness that exists in Fig.~(\ref{circular_eigen}) is due to the toggling back and forth between edges states of similar energy from the various Landau levels, uniquely identified in Fig.~(\ref{circular_eigen_ell}) [and not
in Fig.~(\ref{circular_eigen})]. The Landau levels are distinct and clear, except now they carry on indefinitely, to the left in  Fig.~(\ref{circular_eigen_ell}a), and to the right in Fig.~(\ref{circular_eigen_ell}b). As already stated and further corroborated below, the increase in energy is due to the confinement of the edge. The `toggling' is evident if one examines the states at a
dimensionless energy value like $\approx 1400$ in Fig.~(\ref{circular_eigen_ell}a). The next energy level that would be placed in Fig.~(\ref{circular_eigen})
would come from one of the two branches either emanating from the first or the second Landau level, and these would go back and forth. At a 
higher value, say $\approx 2000$, toggling would occur between three branches, and so on, leading to increasing ``raggedness'' in 
Fig.~(\ref{circular_eigen}) as the energy goes up, which is precisely what we observed.
\begin{figure}[H]
\begin{center}
\includegraphics[scale=0.4]{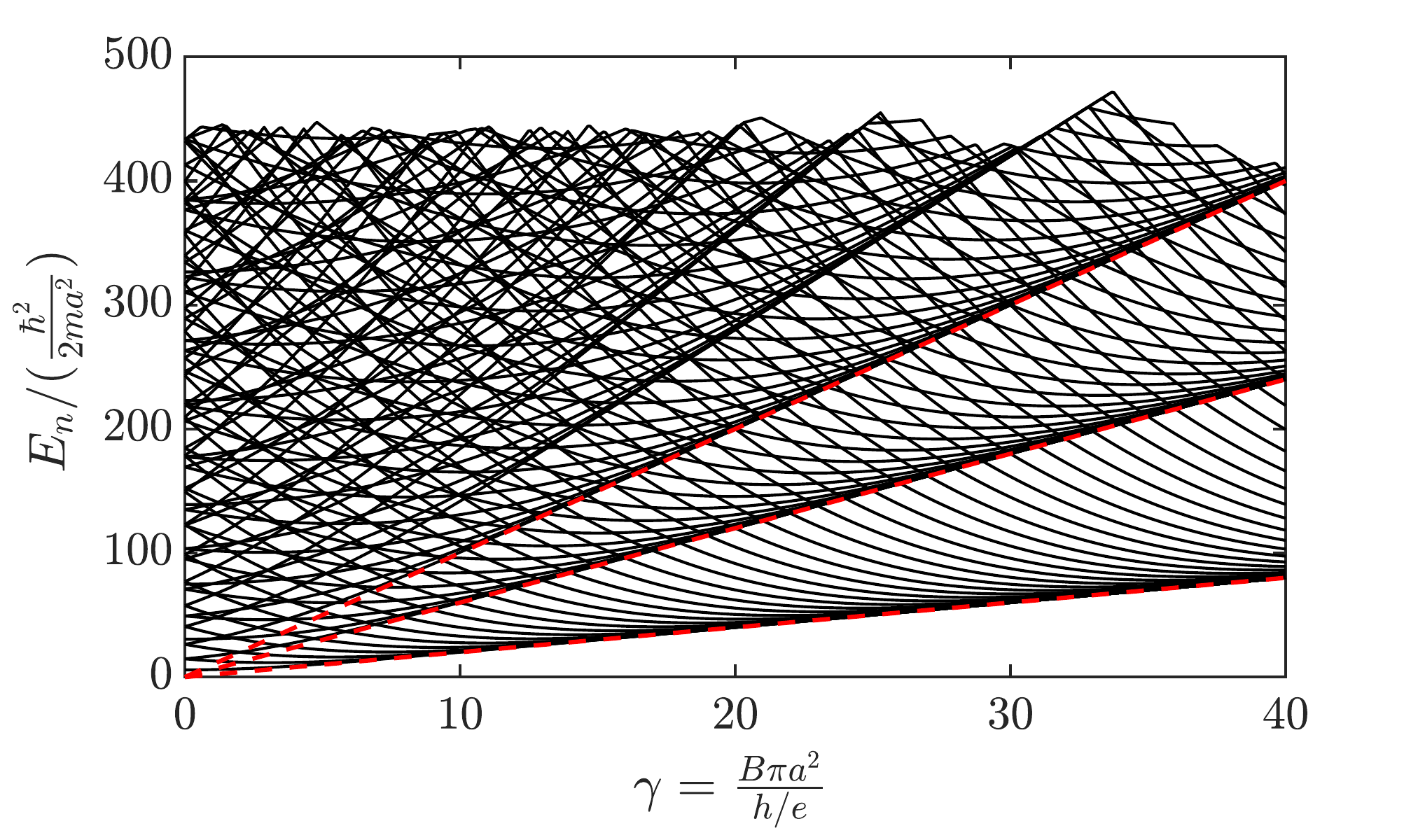}
\end{center}
\caption{The first 100 levels of the Fock-Darwin spectrum for the circular well. Also shown (with red dashed lines) are the free Landau levels. In this case, in contrast
to that shown in Fig.~(\ref{fock_darwin}) with parabolic confinement, the levels ``condense'' to these free Landau levels as the magnetic field increases, as illustrated
by Lent (his Fig.~1).\cite{lent91} These condensed bands are the (quasi) degenerate plateaus seen for fixed $\gamma$ in Fig.~\ref{circular_eigen}. The discrete levels that fill the gaps between condensed bands are the edge states that come after plateaus, also seen in Fig.~\ref{circular_eigen}.}
\label{fock_darwin1}
\end{figure}
Finally, we also include a Fock-Darwin spectrum, as in Lent,\cite{lent91} shown in Fig.~(\ref{fock_darwin1}), where the so-called ``condensation'' of levels occurs.\cite{robnik86}
As $\gamma$ increases, more and more levels coalesce as the free space Landau levels become more applicable since the wave functions become more compact and away from the edge. The thick dashed red lines indicate the expected Landau levels indicative of free space. To the far left, with the exception of $\ell = 0$ the eigenvalues come in pairs, corresponding to positive and negative $\ell$ which are degenerate in the absence of a magnetic field.
Such condensation (seen in the far right on this plot) is absent with parabolic confinement, which is presumably why these are called `Landau' levels and not `Fock' levels.

Fig.~\ref{fock_darwin1} is also illustrative of the toggling phenomenon we have just discussed. Consider $\gamma>30$. For fixed $\gamma$, if we move up the plot in energy (vertically), we go from a condensed band of (quasi) degenenerate states to a region of non-degenerate edge states that all belong to the lowest (confined) Landau level (the lowest band in Fig.~\ref{circular_eigen_ell}). Still increasing in energy, we reach another degenerate set before reaching non-degenerate edge energies once more. But now, edge energies from both $n_L=0$ and $n_L=1$ fill the energy gap between degenerate bands. This pattern repeats as we increase in energy, with the $n^{\text{th}}$ gap containing edge states from all confined Landau levels with $n_L = n-1$ and lower.
This phenomenon has been referred to as `Level Crossing' \cite{rontani99} in the context of Fock-Darwin states. The ordering of these crossed levels become less smooth as we go to higher level gaps. Level crossing is conceptually identical to the edge state toggling that we discussed earlier. 

\subsection{\black{Probability Densities}}
\begin{figure}[H]
\begin{center}
\includegraphics[scale=0.55]{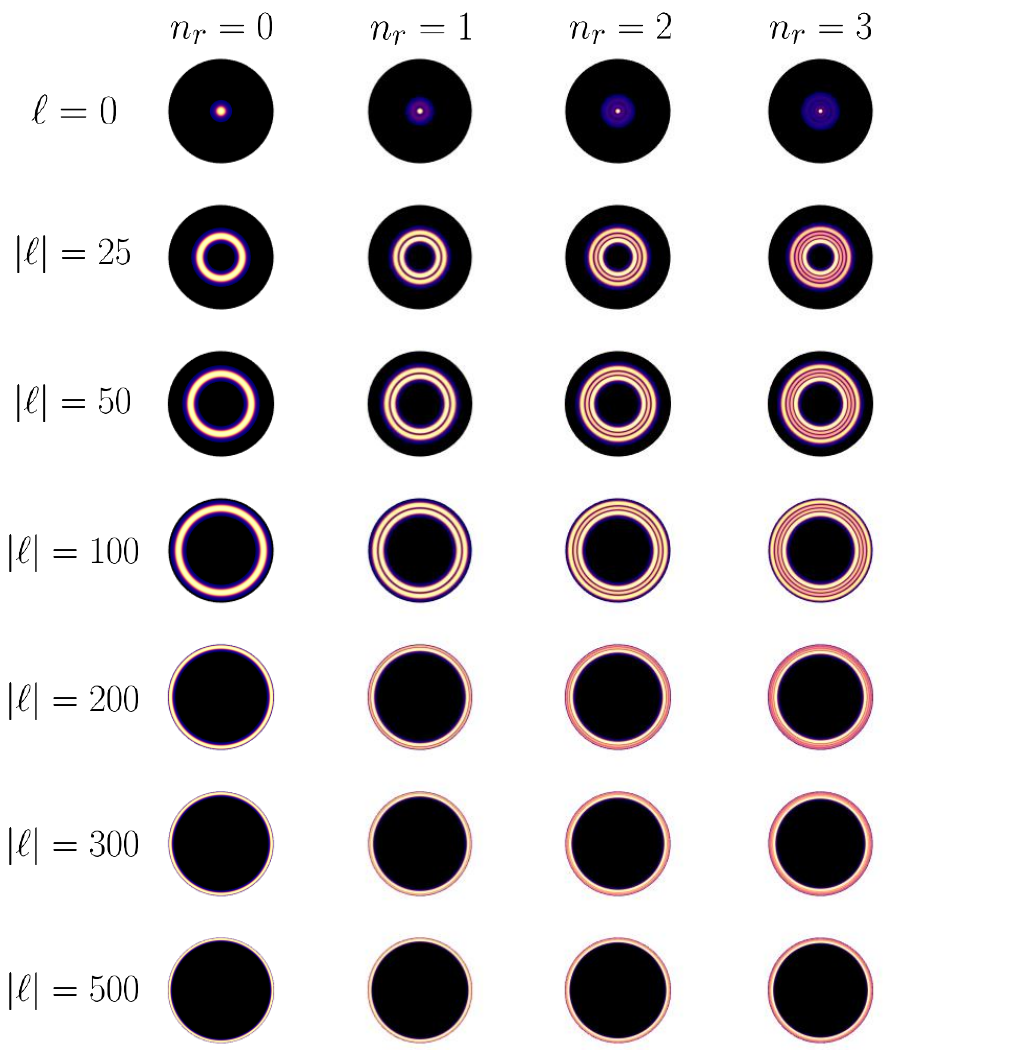}
\end{center}
\caption{Contour plots of $|\Psi_{n_r,\ell}(r,\phi)|^2$ for various values of $n_r$ and $\ell$, for $\gamma = 150$.
The number of nodes increases with increasing $n_r$, more clearly seen for low values of $|\ell|$, and the radius of
the maximal amplitude increases with increasing $|\ell|$. The notion that large $|\ell|$ states are ``edge'' states is clear from this plot, as the entire
probability density resides on the circumference for these states.
}
\label{prob_density1}
\end{figure}

\begin{figure}[H]
\begin{center}
\includegraphics[scale=0.5]{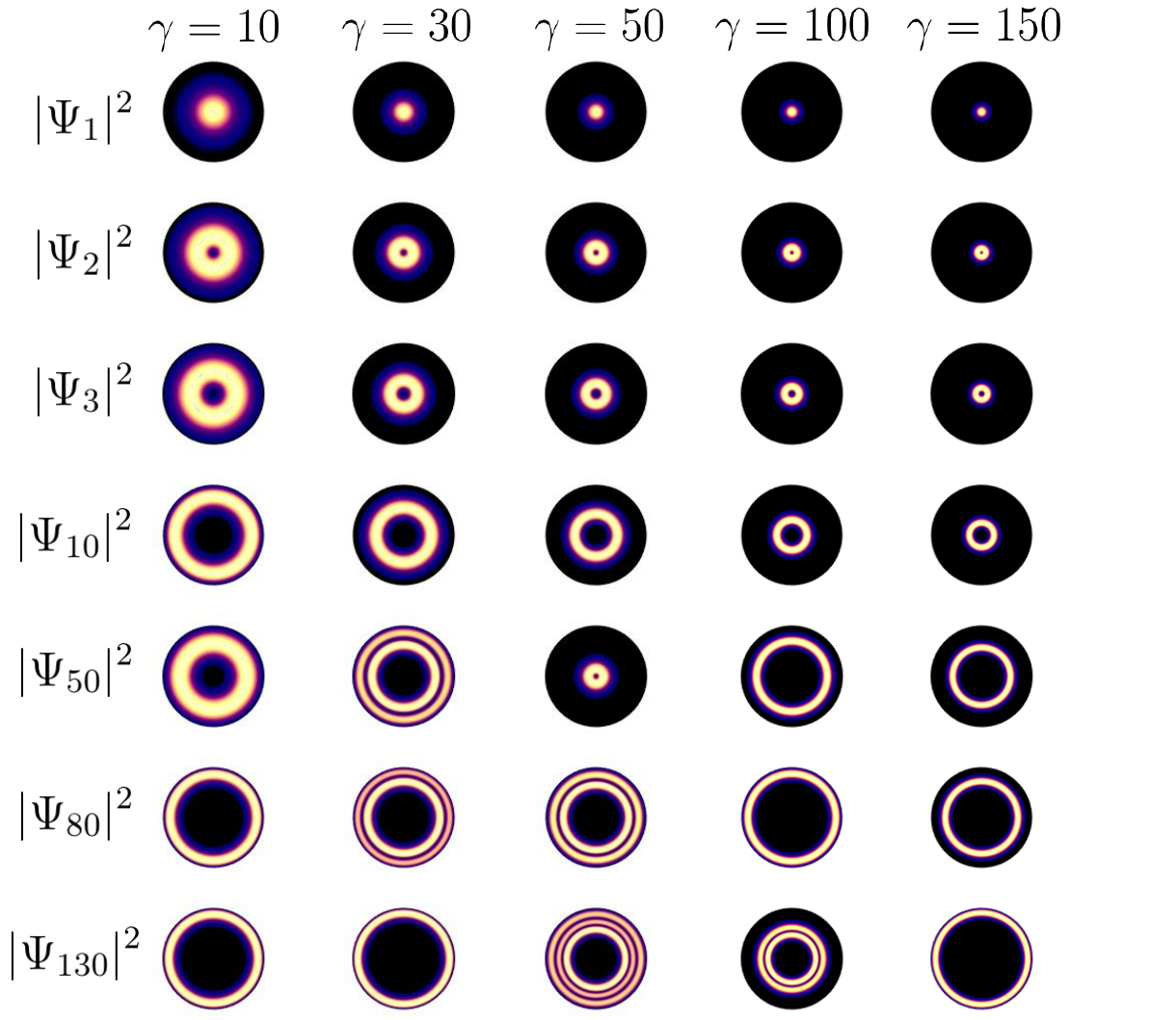}
\end{center}
\caption{Contour plots of $|\Psi_{n}(r,\phi)|^2$ as a function of eigenvalue number $n$ for various values of $\gamma$, as indicated at the top of each column.
Trends similar to those shown in Fig.~(\ref{prob_density1}) are evident, but in addition one can note changes in $\ell$ and $n_r$ quantum numbers. For example,
$|\Psi_{50}|^2$ for $\gamma = 50$ has $n_r = 0$ and $\ell = +1$, and is located at the start of the 2nd plateau visible in Fig.~(\ref{circular_eigen}). It is identical
to $|\Psi_2(r,\phi)|^2$ for $\gamma = 50$ displayed here in the 2nd row. The wave function $\Psi_2(r,\phi)$ has quantum numbers $n_r=0$ and $\ell = -1$ 
(recall that $|\Psi|^2$ was insensitive to
the sign of $\ell$). Note that this plot is orderly and symmetric because we have artificially broken the remaining quasi-degeneracy apparent in Figs.~(\ref{circular_eigen})
and (\ref{circular_eigen_ell}) with a {\it very} shallow parabolic trap, centred at the middle of the cylinder. This shallow trap changes nothing that is visible to the eye,
but it does remove the remaining quasi-degeneracy so that a proper ordering is established. See the text and the next section for further explanation of this additional
potential, whose sole purpose is to remove the degeneracy.}
\label{prob_density1b}
\end{figure}

We have been discussing edge states, and using the parameter $r_\ell \equiv l_B\sqrt{2|\ell|}$ (recall $l_B  \equiv \sqrt{\hbar/(eB)}$ is the magnetic length) to
indicate the location of the wave function. Of course $r_\ell$ is the radius of the maximum of the wave function (from some specified origin) in free space only, so here
we examine the actual probability densities. Recall that the probability density is gauge invariant, except in cases of degenerate wave functions. Here the degeneracy is
removed because of the confining cylinder, but as already discussed, a `practical' degeneracy remains, so the statement about gauge invariance is no longer true,\cite{swenson89}
as we shall see below and in the next section.\\
In Fig.~(\ref{prob_density1}) we show contours of $|\Psi_{n_r,\ell}(r,\phi)|^2$ for $\gamma=150$ for various values of $n_r$ and $\ell$.
These indicate that the wave function amplitude moves out from the centre as $|\ell|$ increases, as noted analytically for the free space result. Moreover, it is
clear that the number of nodes increases as $n_r$ increases, and in fact this quantum number tells us the number of nodes.
 We should note that Fig.~(\ref{prob_density1}) looks very ``orderly'' when it comes to the progression of the number of nodes (as one moves to the right, increasing $n_r$) and the radius of maximum amplitude (as one moves down, increasing $|\ell|$). This is because we have complete control of the quantum numbers, as stressed with respect to 
Fig.~(\ref{circular_eigen_ell}). This scenario is different when we simply order all the states according to their energies [this was done to produce Fig.~(\ref{circular_eigen})].
This will make the progression of states somewhat disordered, {\it for a reason entirely different} than the disorder already noted in connection with Fig.~(\ref{circular_eigen}). {\it That} disorder was visible to the eye. Now we are referring to the quasi-degenerate states in the plateau regions, where the disorder {\it is
not visible, but nonetheless exists at a level lower than $10^{-16}$} (so it is invisible even to the computer, using double precision accuracy). For this reason,
even in the plateau regions in Fig.~(\ref{circular_eigen}), the states may be ``out of order,'' and a state with a higher $|\ell|$ may be ranked
lower (according to energy) compared with a lower $|\ell|$ state. To avoid this we artificially include a {\it very shallow} parabolic trap, whose sole purpose is to break
the remaining quasi-degeneracy, so that the progression of states is orderly. The resulting contour plots are shown in Fig.~(\ref{prob_density1b}).

 We will discuss in more detail
the introduction of this ``degeneracy-breaking'' potential in the next section. Including it as we do here does not change any of the results concerning energy
that are visible to the eye; it does order the quasi-degenerate energy levels at the $10^{-10}$ level, so the computer ``knows'' which states are supposed to come first, and does
not resort to a somewhat random linear combination, as happens when there is a degenerate subspace of solutions.
 
\subsection{\black{Probability Current in a Circular Dot}}

In the Integer Quantum Hall Effect, edge states (eigenstates localized along a boundary) are often cited as a key ingredient to explain the quantization of hall conductance.~\cite{qheTong} Classically speaking, the idea is that if an electron encounters a boundary while undergoing a cyclotron orbit, it will reflect off the wall and once again try to complete its orbit. This causes the electron to undergo 'skipping' orbits along the boundary, resulting in chiral edge currents along the boundary of the sample.
One would expect that edge states would contribute non-zero current to the system while states in the bulk would not contribute any current. Lent~\cite{lent91} illustrated such currents, and since the matrix mechanics technique allows for easy numerically exact calculations of this property, we will illustrate them here as well. Emphasis will be placed on current densities corresponding to eigenstates of the LLL.

The probability current density $\mathbf{J}$ of an electron in an eigenstate $\Psi$ immersed in a magnetic field represented by a gauge choice $\mathbf{A}$ is given by~\cite{landau77,wysin}
\begin{equation}
    \mathbf{J}=\frac{1}{2m}\left[\Psi^*\mathbf{p}\Psi-\Psi\mathbf{p}\Psi^*+2e|\Psi|^2\mathbf{A}\right].
    \label{current1}
\end{equation}
With this equation and Eq.~(\ref{psi_transform}), it can easily be shown that $\mbf{J}$ is a \textit{gauge-independent} quantity for non-degenerate eigenstates. However, the free space infinite degeneracy present in this problem relates the eigenstates of the two gauges in a not so trivial manner, as mentioned earlier. Consequently, the expressions for the current density in each gauge radically differ. We will return to this issue when we investigate probability currents in a square quantum dot. Here, we present an analysis of the behaviour of the current density in the symmetric gauge.

Since we are working in two dimensions, $\mbf{J}$ is a \textit{surface} current density with dimensions $IL^{-1}$. Using the symmetric gauge $\vec{\bf A}_s$, the probability current density associated with an eigenstate $\Psi_{n_r,\ell}(r,\phi)=\frac{e^{i\ell\phi}}{\sqrt{2\pi}}\psi_{n_r,\ell}(r)$ is, in polar coordinates, 
\begin{equation}
    \mathbf{J}_{n_r,\ell}(r)=\frac{|\psi_{n_r,\ell}(r)|^2}{2\pi}\left[\frac{\hbar \ell}{mr}+\frac{1}{2}\omega_cr\right]\boldsymbol{\hat{\phi}}=\frac{|\psi_{n_r,\ell}(r)|^2}{2\pi}\frac{\Lambda_z(r)}{mr}
    \boldsymbol{\hat{\phi}},
    \label{current2}
\end{equation}
where $\Lambda_z(r)=\hbar \ell+\frac{1}{2}m\omega_cr^2$ is the kinetic angular momentum in the $\hat{\boldsymbol{z}}$ direction. 
For further study, we focus on the current density in the LLL with $n_L=0$. This corresponds to states with $n_r=0$ and $\ell\leq 0$.

\begin{figure}[H]
\begin{center}
\includegraphics[scale=0.3]{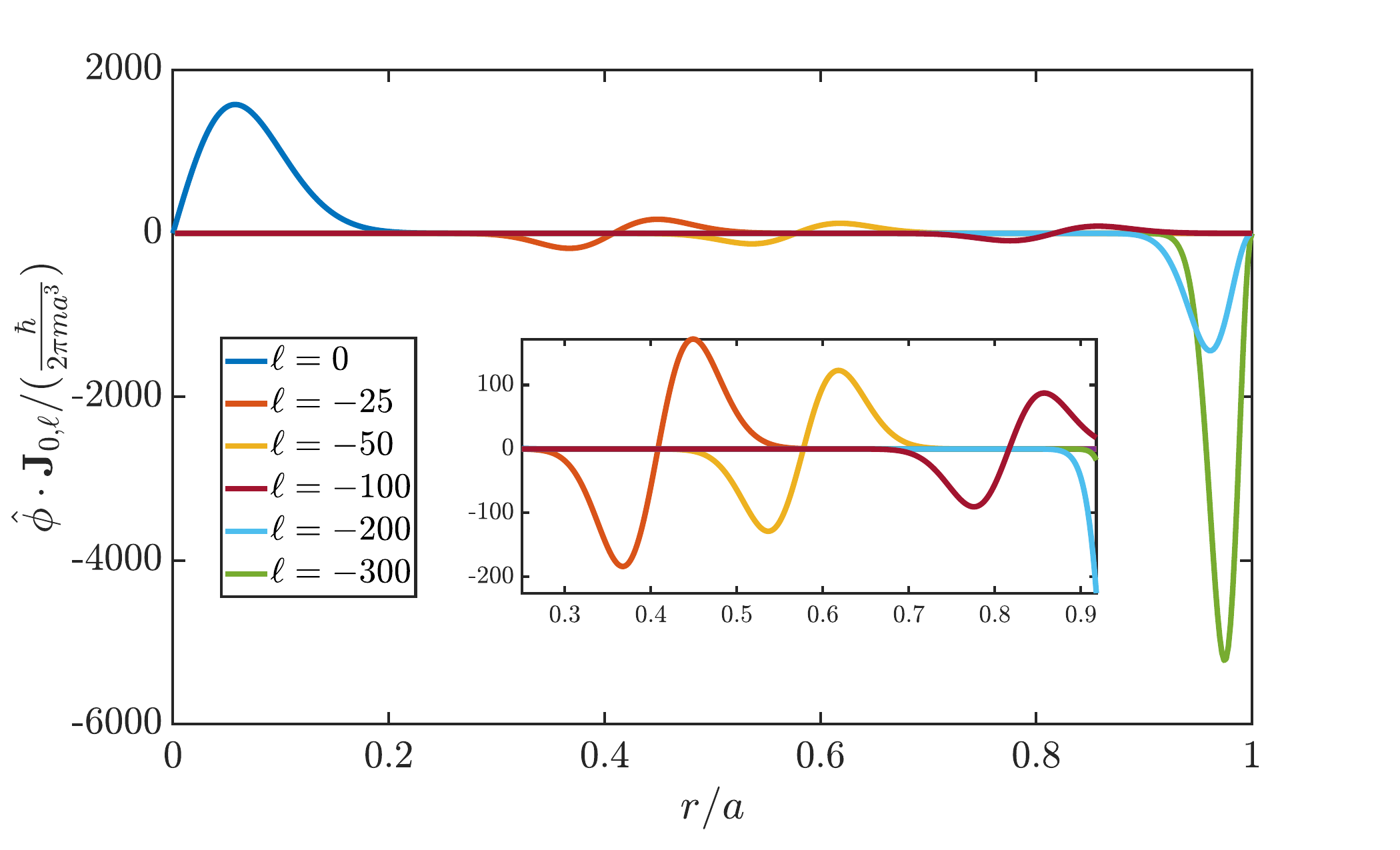}
\end{center}
\caption{Dimensionless probability current density magnitude in the confined LLL plotted vs $r/a$ for various $\ell\leq 0$ and fixed $\gamma=150$. Convergence was achieved with $N=200$. Note how the $\ell=0$ state has current 
density concentrated near the origin, while other (negative) $\ell$ states have current density located (naturally) where their probability density dominates, but with an inflection point [as predicted from Eq.~(\ref{current2})
right at $r/a = r_\ell/a = \sqrt{|\ell|/\gamma}$, {\it as long as the probability density is well away from the edge.} These states are representative of bulk currents and behave according to Eq.~(\ref{LLL_curr}). The larger $|\ell|$ states are representative of edge currents, for which the current density is concentrated near the edge and flows exclusively in the negative $\boldsymbol{\hat{\phi}}$ direction.}
\label{prob_current2}
\end{figure}

To predict the behaviour of the LLL current density in the bulk of our dot, we construct an expression for $\mbf{J}$ in \textit{free space}. It readily follows from Eq.~(\ref{current2}) that the current density in the LLL is given by:
\begin{equation}
    \mbf{J}^{(FS)}_{0,\ell}(r) = \frac{1}{2\pi}\left(\frac{r}{\sqrt{2}l_B}\right)^{2|\ell|}\frac{e^{-\frac{r^2}{2l_B^2}}}{l_B^2|\ell|!}\left[\frac{\hbar \ell}{mr}+\frac{1}{2}\omega_cr\right]\boldsymbol{\hat{\phi}} \ \ \ \ \ \ell \leq 0. \label{LLL_curr}
\end{equation} 
We have added a superscript `$(FS)$' to emphasize that this expression holds in free space only. Note that this expression is still useful to understand the behaviour of bulk states in a quantum dot only because we have located the origin for the Symmetric gauge precisely at the centre of the quantum dot. This expression for the probability current will only apply as long as the wave functions remain essentially zero at the edge of the quantum dot, which will be true as long as $|\ell|$ is small, or more precisely, $r_{\ell} << a$. This is a way of defining what a "bulk" state is for this system.
One can verify that $\mbf{J}^{(FS)}_{0,\ell}$ is localized near $r\approx r_\ell=l_B\sqrt{2|\ell|}$ but vanishes precisely at $r=r_\ell$. With the exception of $\ell = 0$, it can also be shown that $r=r_\ell$ is also a point of inflection for the current density. We expect this behaviour for lower $\ell$ bulk states in our confined system, which we plot along with edge states in Fig.~(\ref{prob_current2}).

\begin{figure}[H]
\centering
\includegraphics[scale=0.4]{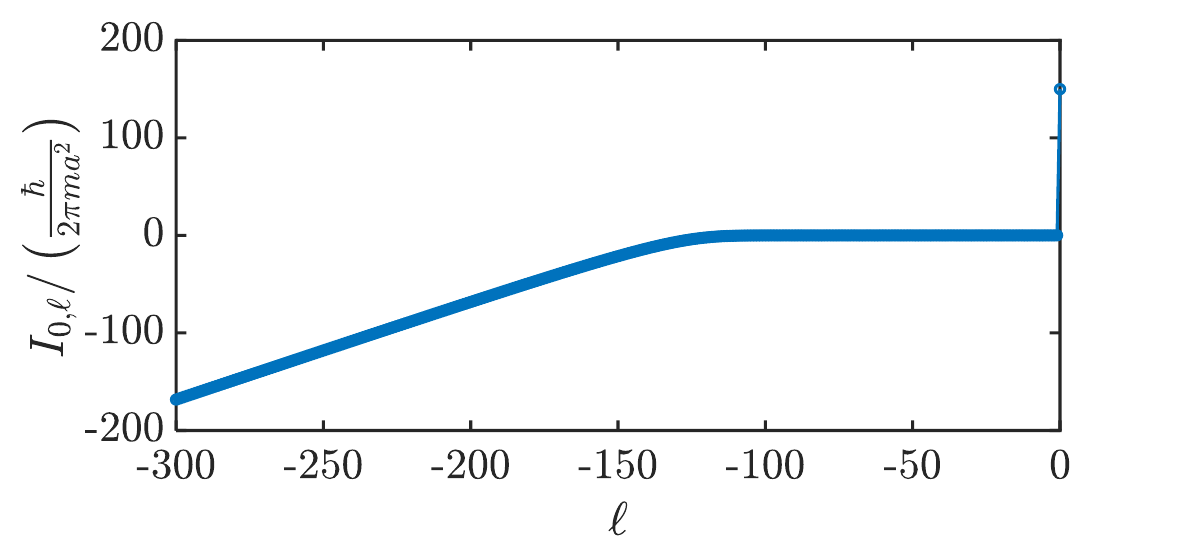}
\caption{Dimensionless probability current (Eq.~\ref{current4}) vs angular momentum quantum number $\ell$ for $\gamma=$150. With the exception of $\ell=0$, which contributes precisely $\gamma$ current, bulk states contribute zero current to the system. This is in accordance to what we calculated for the free space current in Eq.~(\ref{current10}). For large values of $|\ell|$, we see that edge states contribute nonzero current that flows in the negative $\boldsymbol{\hat{\phi}}$ direction. As $|\ell|$ increases further, the edge states become more localized at the boundary and contribute increasingly more current to the system.}
\label{tot_curr_LLL}
\end{figure}

Through the inset in Fig.~(\ref{prob_current2}), we see that the total area bounded by the current density of the bulk states (aside from $\ell=0$) appears to vanish (equal areas of positive and negative current density cancel one another) due to the inflection point present at $r=r_\ell$. This is not true once $r_\ell$ approaches (and mathematically) exceeds the radius $a$ of the dot, and so edge current densities bound non-zero area. The bounded area in Fig.~\ref{prob_current2} is the net \textit{current} flow in the $\boldsymbol{\hat{\phi}}$ direction and as, mentioned earlier, we would expect that this current is non-zero only near the edge. The notion of non-zero edge currents and the results of Fig.~\ref{prob_current2} appear to be in qualitative agreement, but this agreement can easily be quantified. To obtain the current from a surface current density, we integrate it with respect to a scalar differential `strip' orthogonal to the direction of flow. Thus, the state current $\mbf{I}_{n_r,\ell}$ of a given eigenstate in our confined system is given by
\begin{equation}
    \mbf{I}_{n_r,\ell}=\int_{0}^{a} \mbf{J}_{n_r,\ell}(r)dr.
    \label{current4}
\end{equation}
Since the current densities are quite localized, we can accurately estimate what to expect for a state far from the edge using the free space expression for current density 
Eq.~(\ref{LLL_curr}). The free space current for an LLL state is calculated via:
\begin{equation}
    \mbf{I}_{0,\ell}^{(FS)} = \int_{0}^{\infty}dr\mbf{J}^{(FS)}_{0,\ell}(r) = \int_{0}^{\infty}dr\frac{1}{2\pi}\left(\frac{r}{\sqrt{2}l_B}\right)^{2|\ell|}\frac{e^{-\frac{r^2}{2l_B^2}}}{l_B^2|\ell|!}\left[\frac{\hbar \ell}{mr}+\frac{1}{2}\omega_cr\right]\boldsymbol{\hat{\phi}}, \ \ \ \ \ \ell \leq 0.
    \label{state_curr1}
\end{equation}
Defining $I_{0,\ell}^{(FS)}\equiv\mbf{I}_{0,\ell}^{(FS)}\cdot\boldsymbol{\hat{\phi}}$ and using current units ($\hbar/(2\pi ma^2$) for convenience, we evaluate the integral in Eq.~(\ref{state_curr1}) as follows:
\begin{equation}
\begin{split}
    I^{(FS)}_{0,\ell}/\frac{\hbar}{2\pi ma^2}\equiv i^{(FS)}_{0,\ell}&=\int^{\infty}_0\frac{2\gamma^{|\ell|+1}x^{2|\ell|}}{|\ell|!}e^{-\gamma x^2}\left[\frac{\ell}{x}+\gamma x\right]dx\\[5pt]
    &=\frac{2\gamma^{|\ell|+1}\ell}{|\ell|!}\int^{\infty}_0x^{2|\ell|-1}e^{-\gamma x^2}dx+\gamma\underbrace{\int_0^{\infty}\frac{2\gamma^{|\ell|+1}x^{2|\ell|}}{|\ell|!}e^{-\gamma x^2}xdx}_{=1}\\
    &=\gamma \frac{\ell}{|\ell|!}\underbrace{\left(\int^{\infty}_0x^{|\ell|-1}e^{-x}dx\right)}_{=\Gamma(|\ell|)=(|\ell|-1)!}+\gamma \\
    &=\gamma [\text{sign}(\ell)+1]\\[5pt]
   \Rightarrow i^{(FS)}_{0,\ell}  &=
   \begin{cases}
    2\gamma \ &\text{if} \ \ell>0\\
    \gamma \ &\text{if} \ \ell=0\\
    0 \ &\text{if} \ \ell<0.\\
    \end{cases}
    \label{current9}
\end{split}
\end{equation}
Therefore, the state current magnitude in free space is given by:
\begin{equation}
I^{(FS)}_{0,\ell}
=\begin{cases}
    f_c \ &\text{if} \ \ell>0\\
    f_c/2 \ &\text{if} \ \ell=0\\
    0 \ &\text{if} \ \ell<0,\\
    \end{cases}
    \label{current10}
\end{equation}
where $f_c\equiv\omega_c/(2\pi)$. Thus, in our confined system, we expect the bulk states in the LLL to contribute zero current, with the exception of the $\ell=0$ state. Results are presented for positive $\ell$ (but $n_r=0$) states as well. These states exist outside the LLL, but since we did not have to restrict the sign of $\ell$ to proceed with our calculation, we show this result for completeness. Note that these positive $\ell$ states contribute current that flows in the positive $\boldsymbol{\hat{\phi}}$ direction. In contrast to the negative $\ell$ states, the positive $\ell$ states all circulate in the direction expected classically (through the Lorentz force), but are energetically unfavourable.
Numerical results for probability current in our confined system are shown in Fig.~(\ref{tot_curr_LLL}).

\begin{figure}[H]
    \centering
    \includegraphics[scale=0.3]{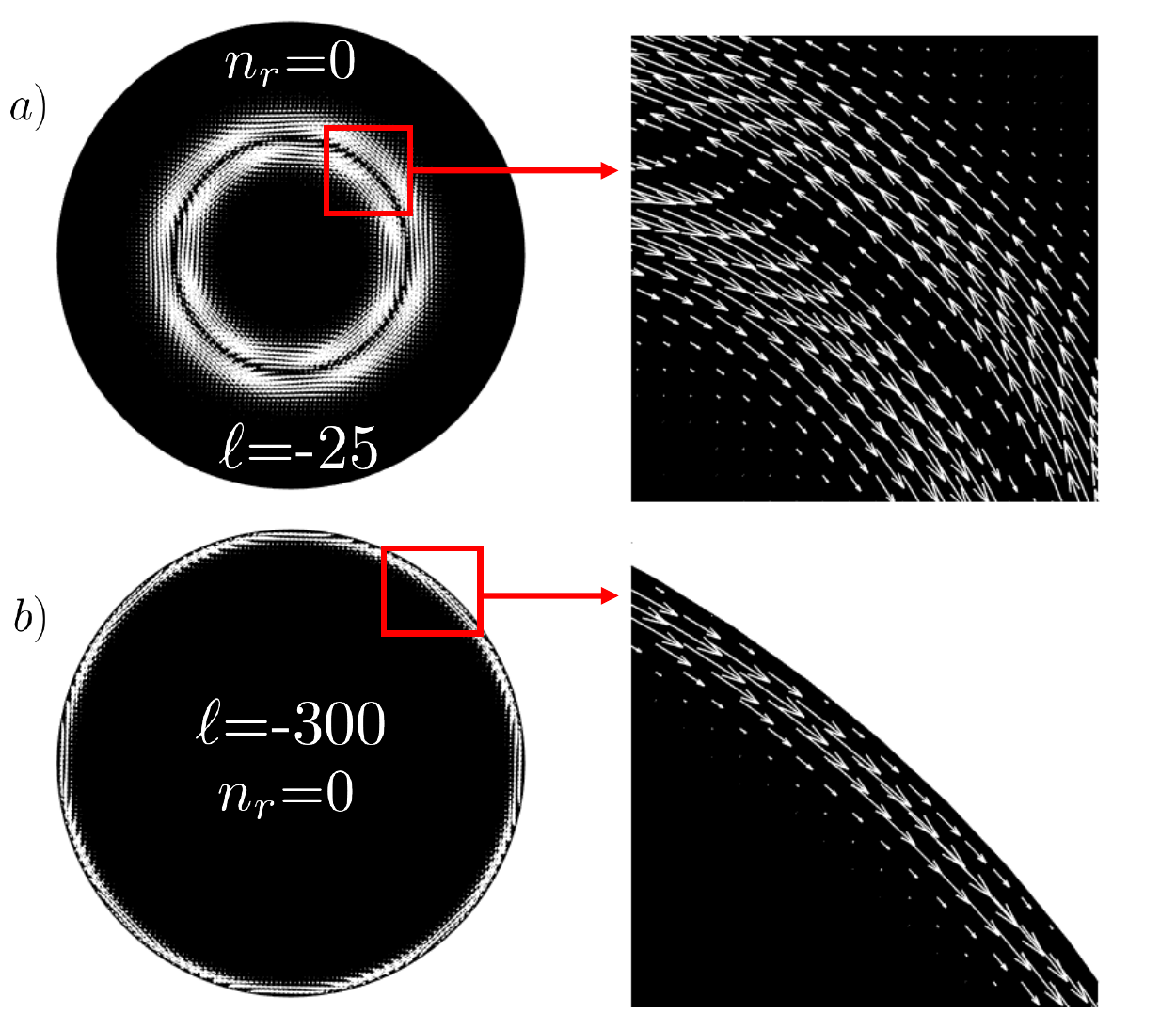}
    \caption{A vector field plot of the probability current for a ``bulk'' state in (a) ($\ell = -25$, $n_r = 0$) and an edge state in (b) ($\ell=-300$, $n_r = 0$). 
    In both cases we used $\gamma=100$, and convergence was attained with  a matrix size of $N=200$. The expanded portions show what should 
    already be clear from Fig.~(\ref{prob_current2}). Although the edge state in b) contributes non-zero current along the boundary, it does not exhibit the classically expected skipping orbit trajectory. 
    }
    \label{circ_vec_field}
\end{figure}

Finally, we show a vector plot of the current for an LLL ``bulk'' state in (a) ($\ell = -25$) and an LLL edge state in (b)
($\ell = -300$) in Fig.~(\ref{circ_vec_field}) , with the plots on the right showing the same result for the region in  the red square box expanded in more detail.

Our numerical results are in qualitative agreement with those shown by Lent,\cite{lent91}
although he utilized smaller values of $\gamma$ (his $\beta \equiv 2\gamma$), and he showed pictorial vector plots for fairly small magnitudes of $\ell$ (which he calls $m$). He also provides a nice description in terms of classical orbits, which we will not repeat here; the reader is referred to Ref.~[\onlinecite{lent91}] for this description.

\vspace{0.25in}
\section{\black{Landau Levels in a Square Quantum Dot}}
\label{square}
Relatively little work has been done to date for a square geometry.
In this section we present results in both the Symmetric and Landau gauges, and illustrate the difficulties encountered for large fields (or samples) due to the practical degeneracy that remains in these cases. Since the square geometry is rather difficult in either gauge, we thought it worthwhile to provide results for the same geometry in both gauges. That this will work illustrates the remarkable machinery of gauge invariance, as now the gauge choice results in a very non-symmetric-looking potential. In fact as we will shortly see it doesn't work for large values of $G$, for the reasons just described above, but our very slight perturbing confining potential will fix the problem (i.e. lift the degeneracy) so that identical results are achieved in both gauges.

\subsection{\black{Matrix Element Calculation}}

\noindent\textbf{Symmetric Gauge}

The confining potential is defined as
\begin{equation}
    V(x,y)=\begin{cases}
    0 \ &\text{if} \ 0< x < a \ \  {\rm and} \ \ 0<y<a\\
    \infty \ &\text{otherwise}.
    \label{pot_confining_sq}
    \end{cases}
\end{equation}
Here $a$ is the length of a side of the two-dimensional infinite square well representing the square quantum dot. 
Given this potential, a convenient set of normalized basis states is
\begin{equation}
    \phi_{n_x,n_y}(x,y) = \braket{x,y}{n_x,n_y}=\frac{2}{a}\sin{\left(\frac{n_x\pi x}{a}\right)}\sin{\left(\frac{n_y\pi y}{a}\right)},
    \label{basis_sq}
\end{equation}
where $n_x$ and $n_y$ are positive integers. For convenience, we modify the symmetric gauge $\mbf{A}_S$ such that it is centred in the potential well:
\begin{equation}
   \mbf{A}'_{S}=\frac{B}{2}\left[\left(x-\frac{a}{2}\right)\hat{\mbf{y}}-\left(y-\frac{a}{2}\right)\hat{\mbf{x}}\right].
   \label{symm_shifted}
\end{equation}
With this gauge $\mbf{A}'_S$, our Hamiltonian takes the form:
\begin{equation}
    H=\frac{\textbf{p}^2}{2 m_e}-i\frac{\hbar\omega_c}{2}\left[\left(x-\frac{a}{2}\right)\partial_y-\left(y-\frac{a}{2}\right)\partial_x\right]+\frac{1}{8}{m_e}\omega_c^2\left[\left(x-\frac{a}{2}\right)^2+\left(y-\frac{a}{2}\right)^2\right] + V(x,y)
    \label{ham_sq}
\end{equation}
where we use $m_e$ as the mass of the electron to avoid confusion with matrix indices. Following the steps leading to Eq.~(\ref{matrix1})
we then expand the eigenstates of (\ref{ham_sq}), $\ket{\Psi}$, as:
\begin{equation}
    \ket{\Psi}=\sum_{n_x,n_y}c_{n_x,n_y}\ket{n_x,n_y}.
    \label{expand}
\end{equation}
Thus, in energy units of $E_0 \equiv \hbar^2\pi^2/(2{m_e}a^2)$, and using $n\equiv (n_x,n_y)$ and $m \equiv (m_x,m_y)$ for short, our dimensionless matrix elements,
$h_{n,m} \equiv H_{n,m}/E_0$ for the Hamiltonian for this problem are given by\\[3pt]
\begin{equation}
\begin{split}
    h_{n,m}&=\delta_{n_x,m_x}\delta_{n_y,m_y} \left[n_x^2+n_y^2+\frac{G^2}{2}\left(\frac{1}{3}-\frac{1}{(\pi n_x)^2}-\frac{1}{(\pi n_y)^2}\right)\right]\\[5pt]
        &+2 \left(\frac{G}{\pi}\right)^2 \Bigl( (1-\delta_{n_x,m_x})\delta_{n_y,m_y}  g_e(n_x,m_x)    + (1-\delta_{n_y,m_y})\delta_{n_x,m_x}  g_e(n_y,m_y) \Bigr)\\[5pt]
    &-i\frac{16G} {\pi^3}(1-\delta_{n_y,m_y})(1-\delta_{n_x,m_x})\left[\left(g_o(n_y,m_y)f(n_x,m_x)-g_o(n_x,m_x)f(n_y,m_y)\right)\right]
\end{split}
\label{ham_sq_dim}
\end{equation}
with
\begin{equation}
    g_e(n,m)
    =\begin{cases}
    0 \ &\text{if} \ n+m=\text{odd}\\
    \frac{1}{(n-m)^2}-\frac{1}{(n+m)^2} \ &\text{if} \ n+m=\text{even}
    \end{cases}
    \label{ge_def}
\end{equation}
\begin{equation}
    g_o(n,m)
    =\begin{cases}
    \frac{1}{(n-m)^2}-\frac{1}{(n+m)^2} \ &\text{if} \ n+m=\text{odd}\\
    0 \ &\text{if} \ n+m=\text{even}
    \end{cases}
        \label{go_def}
\end{equation}
\begin{equation}
    f_o(n,m)
    =\begin{cases}
     \frac{nm}{n^2-m^2} \ &\text{if} \ n+m=\text{odd}\\
    0 \ &\text{if} \ n+m=\text{even}
    \end{cases}
        \label{fo_def}
\end{equation}
where 
\begin{equation}
    G=\frac{Ba^2}{h/e} \equiv \frac{\Phi_{\phantom{a}}} {\Phi_0}
\end{equation}
as defined earlier, and the subscripts `e' and `o' in the definitions, Eqs.~(\ref{ge_def}, \ref{go_def}, \ref{fo_def}) serve to remind us that these are respectively 
non-zero for even or odd sums of the indices only.\\
\\
\noindent\textbf{Landau Gauge}\\[3pt]
We now use the (square well-centred) Landau gauge. This is given by
\begin{equation}
    \mbf{A}'_L=B\left(x-\frac{a}{2}\right)\hat{\mbf{y}}.
    \label{landau_shifted}
\end{equation}
With this gauge and the confining potential defined in Eq.~(\ref{pot_confining_sq}), 
the Hamiltonian is
\begin{equation}
    H=\frac{\textbf{p}^2}{2{m_e}}-i\hbar\omega_c\left(x-\frac{a}{2}\right)\partial_y+\frac{1}{2}{m_e}\omega_c^2\left(x-\frac{a}{2}\right)^2 +V(x,y).
    \label{ham_landau}
\end{equation}
The Hamiltonian, again given in units of $\hbar^2\pi^2/(2{m_e}a^2)$, becomes
\begin{equation}
\begin{split}
    h_{n,m}&=\delta_{n_x,m_x}\delta_{n_y,m_y} \left[ n_x^2+n_y^2 + \frac{G^2}{3} \left( 1 - \frac{6}{(\pi n_x)^2} \right) \right]\\[5pt]
    &+8 \left(\frac{G}{\pi}\right)^2  (1-\delta_{n_x,m_x})\delta_{n_y,m_y}  g_e(n_x,m_x)  \\[5pt]
    &+i\frac{32G}{\pi^3}(1-\delta_{n_x,m_x})(1-\delta_{n_y,m_y}) g_o(n_x,m_x)f_o(n_y,m_y)
    \label{ham_landau_nm}
\end{split}
\end{equation}
where the functions $g_e$, $g_o$, and $f_o$ were all defined earlier. This is a very different matrix than that generated for the symmetric gauge, and moreover is very
asymmetric in $x$ and $y$. Yet we expect to obtain the same results as in that gauge.

\subsection{\black{Eigenenergy Spectra}}

The eigenenergies calculated in both gauges were found to be identical to numerical precision, so the results below are representative of both gauges. This is expected of course, as energy is an observable quantity and so \textit{must} be gauge-invariant, regardless of how different the matrix representations look in the two gauges. This is because the Hamiltonians in each gauge choice are unitary transformations of one another, as shown in Eq.~(\ref{Ham_transform}).
Note that a lower truncation value is {\it necessarily} used here since matrices with block matrices as elements have to be constructed and diagonalized, which is much more computationally intensive than in the problem with circular confinement. Thus a truncation of $N=100$ for one of the indices, say $n_x$, implies a total truncation size of $N^2$ for both $n_x$ and $n_y$. We will refer only to the block matrix truncation size ($N$) in the following results (so $N=100$ values of $n_x$ requires diagonalization of a $10000 \times 10000$ matrix).

\begin{figure}[H]
\begin{center}
\includegraphics[scale=0.3]{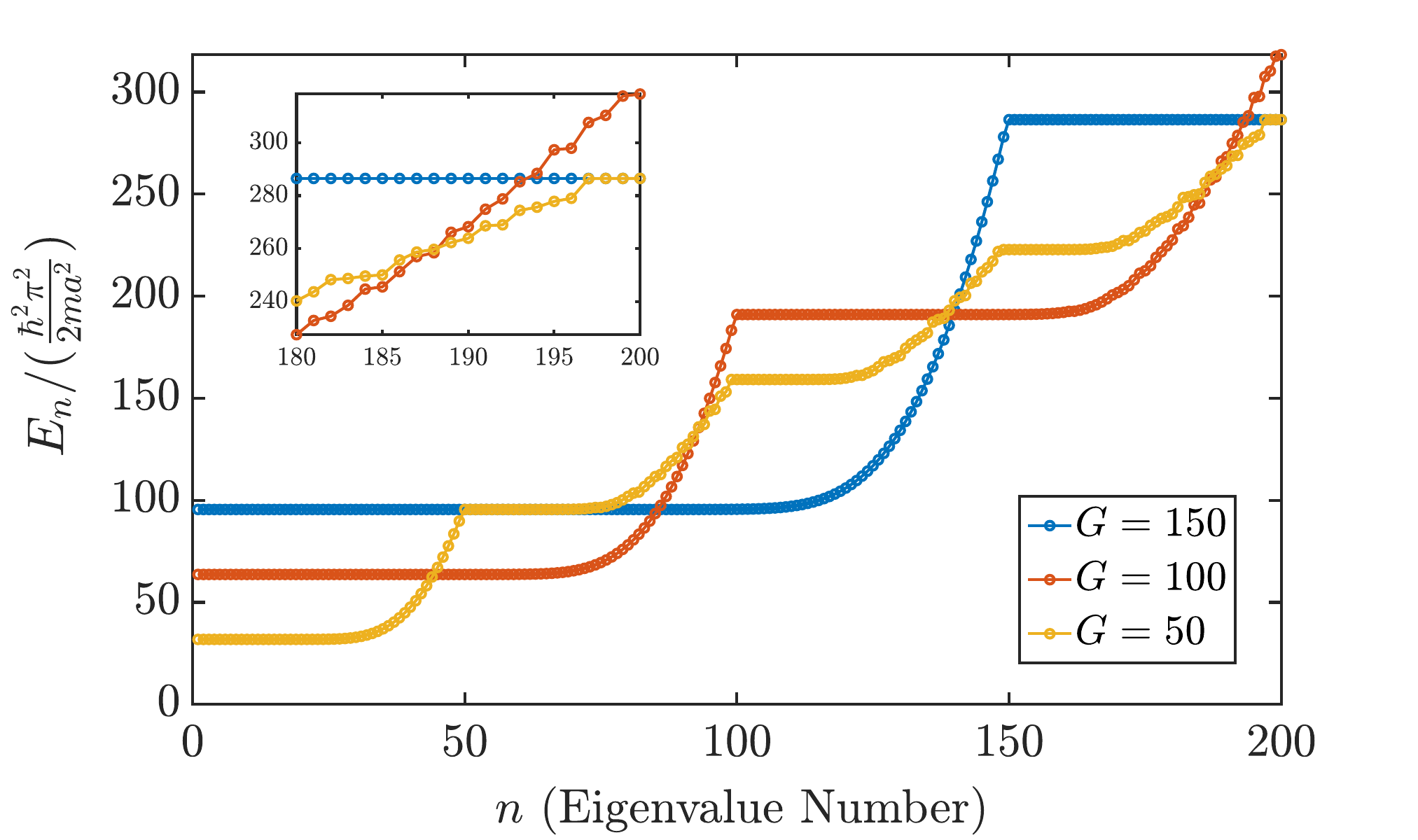}
\end{center}
\caption{Eigenvalues shown in ascending order vs quantum number for $G=50$, $100$, and $150$. This figure is the `square' version of its circular counterpart, Fig.~(\ref{circular_eigen}), and we {\it do not} have the counterpart to Fig.~(\ref{circular_eigen_ell}), which would allow us to sort out here the two quantum numbers that are present in that case. Just like its circular well counterpart, the energies plateau at free space Landau level energies, which are given by $E_{n_L}/\left(\frac{\hbar^2\pi^2}{2ma^2}\right)=\frac{4G}{\pi}\left(n_L+\frac{1}{2}\right)$. Two noticeable features are the plateaus, as in the circular case, especially for large values of $G$, and the number of states present between plateaus, precisely $G$ here, just as it was $\gamma$ for the circular case. Also noticeable is a similar ``raggedness'' for energy values beyond the higher plateaus, as seen clearly for energies occurring past the fourth plateau for $G=50$ in the inset. This is presumably due to the same 'toggling'/'level crossing' phenomenon that we could explicitly identify in the circular case. We used $N=150$ here.}
\label{square_eigen}
\end{figure}

Some of the lowest computed eigenvalues are shown in Fig.~(\ref{square_eigen}) for various values of $G$ as indicated. 
This figure shares many things in common with its circular counterpart, Fig.~(\ref{circular_eigen}), and so it should, since for `bulk-like' states we have argued that they resemble the free space results and hence, do not ``feel'' the edge, i.e. the electrons don't even know if they are confined in a circular or square geometry. 

\begin{figure}[H]
\begin{center}
\includegraphics[scale=0.4]{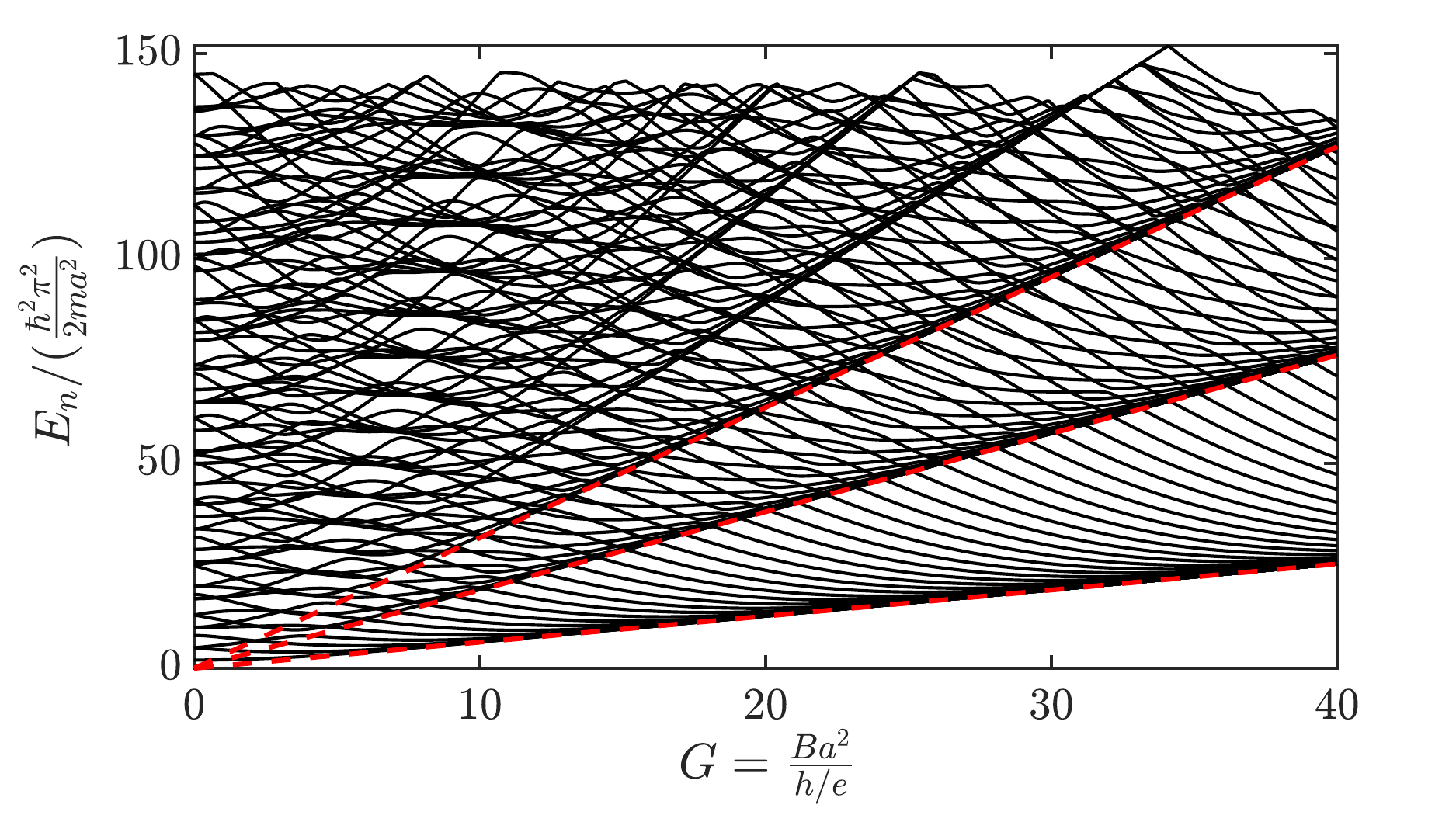}
\end{center}
\caption{The Fock-Darwin spectrum for the two dimensional infinite square well. Here $N=100$.
This plot shares with its counterpart for the circular well, Fig.~(\ref{fock_darwin1}) the idea that for large enough field the levels ``condense'' to a set of degenerate Landau levels (indicated by thicker dashed red lines for the first 3 levels). For weaker fields the levels are fairly disordered, reflecting the more complicated geometry of the square. Level crossing occurs between the condensed energy bands. Note that the zero field energies no longer consistently come in pairs, as the degeneracy pattern for an infinite square well is more complicated than that for an infinite circular well.
}
\label{square_fock_darwin}
\end{figure}

However, the concept of a good quantum number $\ell$ does not exist here, at least not explicitly. So we do not have the benefit of a figure like Fig.~(\ref{circular_eigen_ell}) for the square geometry. One key similarity is the role of the ratio of the total flux to the flux quantum, which gives precisely the number of states between plateaus, regardless of geometry.
In Fig.~(\ref{square_fock_darwin}) we show the Fock-Darwin spectrum for the case of square confinement. As with circular confinement the energy levels eventually ``condense'' and become coalesced into the degenerate Landau levels, as indicated. Both the circular and square confinement show this ``condensation'' phenomenon, whereas the parabolic confinement does not. Like its circular (Fig.~\ref{fock_darwin1}) and parabolic (Fig.~\ref{fock_darwin} in the Appendix) counterparts, 
we see that this spectrum exhibits level crossing between the condensed bands at the free space Landau energies. However, the energies vary in a more complicated manner with increasing flux $G$, in contrast to results seen in both the parabolic and circular dots.\\[1pt]

\subsection{\black{Probability Densities}}
\begin{figure}[H]
\begin{center}
\includegraphics[scale=0.45]{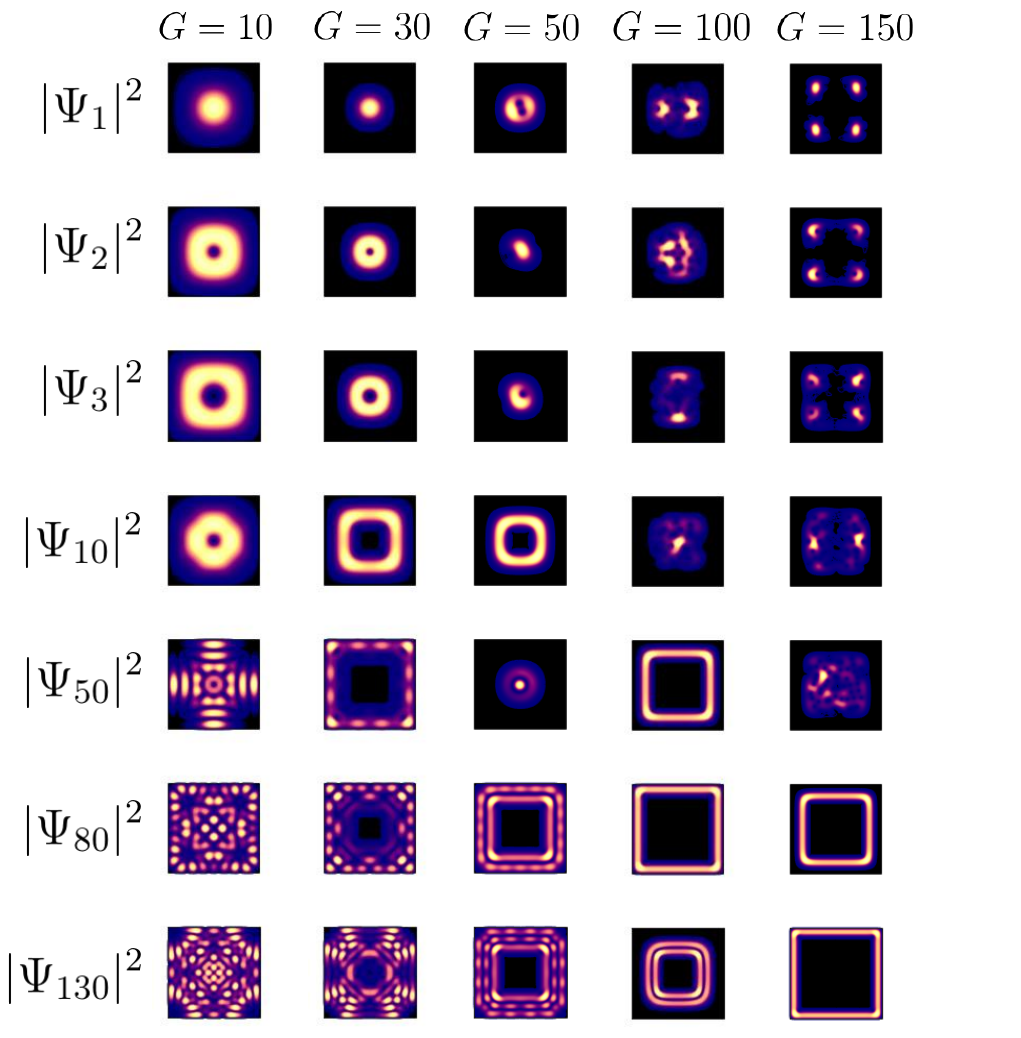}
\end{center}
\caption{Probability density, $|\Psi_{n}(x,y)|^2$ contour plots for various values of eigenvalue number $n$, as ordered by the diagonalization subroutine. Here the confining region is a square, outlined in black. The results only look sensible in the first two columns; the low quantum number results resemble those obtained from the circular geometry, as expected, and then the probability densities become more ``square-like'', reflecting the geometry of the confinement, also as expected. As one increases $G$, however, the results become less understandable, and appear to be wrong. The difficulty, as discussed in the text, is the high degree of degeneracy that remains when large values of $G$ are used. We used $N=150$ for all the square well results.}
\label{prob_density_sq1}
\end{figure}

While the Hamiltonian in both gauges may be related by a unitary operation, in free space, the wave functions of each gauge are not related so trivially. This is due to the infinite degeneracy present in the free space formulation of the problem. In confined space however, this infinite degeneracy is broken. Without this degeneracy, one would expect that Eq.~(\ref{psi_transform}) holds true and that the probability densities should be identical for both gauge choices. However, as we have seen previously with the circular dot, a finite quasi-degeneracy persists to numerical precision. Here, we detail how to get around this issue and find agreement between the probability densities in each gauge.\\
\\
\noindent\textbf{Symmetric Gauge}\\[3pt]
As mentioned previously, for the so-called ``bulk-like'' states we expect the results for the square well to be very similar to those of the circular well, simply because the magnetic field keeps the electron sufficiently contained in the central region of the well so that the electron does not ``know'' the geometry of the confining potential. In Fig.~(\ref{prob_density_sq1}) we show contours of the probability density for a variety of values of G. This plot is to be compared with Fig.~(\ref{prob_density1}), which showed probability densities for the circular well. In that plot, however, we could specify both $n_r$ and $\ell$, whereas here we can only order the plots according to the value of the energy.
Some features of this plot are immediately apparent. First, results with small quantum number and small $G$ are consistent with what we found for circular geometry. This is expected. More excited states become fairly diffuse throughout the square, and retain the symmetry of the square; this is also expected. 
However, as $G$ increases, the results become somewhat irregular. The problem here was already alluded to in the case of the circular 
well: a high degree of ``practical'' degeneracy remains as $G$ increases. We fixed the problem
with little explanation in that case, so now we discuss this issue in more detail.

Because of the remaining ``practical'' degeneracy, the diagonalization routine ends up picking some linear combination of these ``degenerate'' states and ordering them in some fashion. The arbitrary linear combination easily leads to a wave function that does not have the symmetry of the confining potential, as is evident in Fig.~(\ref{prob_density_sq1}). We will see the same problem emerge in the Landau gauge, which {\it does not have} the symmetry of the confinement potential, so it is prudent to emphasize that this 
aspect (having or not having a gauge with the symmetry of the confining potential) is not so important. Here, where the gauge choice 
is the more symmetric one, we still encounter this problem, and it is because of the ``practical'' degeneracy (i.e. energies that are within $<10^{-16}$ of one another) that remains
in the solutions. The `practical'' degeneracy remains because, for large values of $G$, there is enough space in the confined potential for the ``centre'' of the wave
function to be arbitrarily located.

We have resolved this problem by adding an additional perturbative confining potential with the symmetry of the square. Our choice is gauge invariant and is given by:
\begin{equation}
 H' = \frac{1}{2} m_e \omega_0^2 \left[\left(x-\frac{a}{2}\right)^2+\left(y-\frac{a}{2}\right)^2\right],
 \label{hamp}
\end{equation}
where $\omega_0$ is the characteristic frequency to characterize the perturbation potential. It is important to emphasize that this additional
potential is minute: typically $(\omega_0/\omega_c)^2 \approx 10^{-6}$ for $G=50$,
so that the energies are not really affected at any level the eye can detect, but the degeneracy is lifted sufficiently to allow the diagonalization algorithm to
properly and unambiguously order the otherwise quasi-degenerate eigenstates. Most importantly, the centres of all wave functions are now correctly placed at the centre of the square.

\begin{figure}[H]
\begin{center}
\includegraphics[scale=0.45]{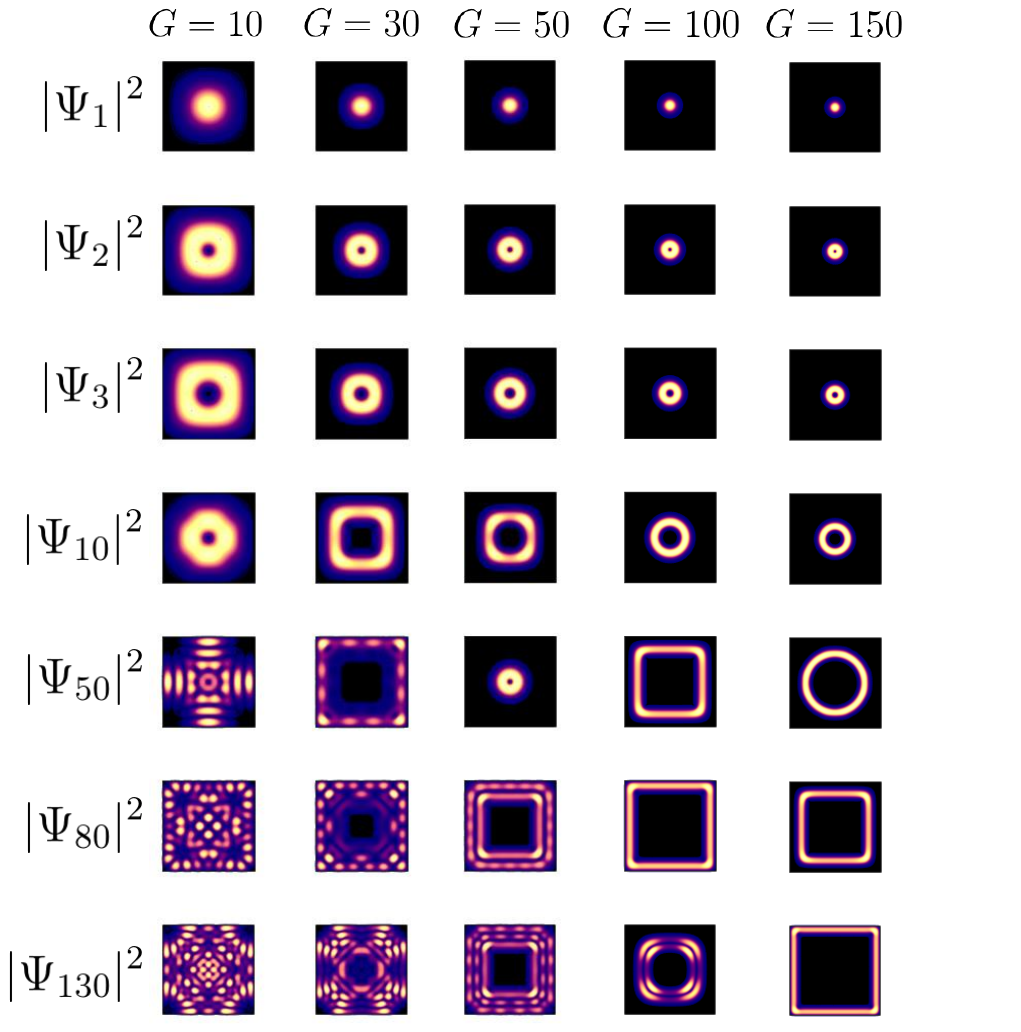}
\end{center}
\caption{Probability density, $|\Psi_{n}(x,y)|^2$ contour plots for the same parameters as in Fig.~(\ref{prob_density_sq1}), but now with an additional confining potential given by
Eq.~(\ref{hamp}) with $\omega_0/\omega_c  =1.3\times 10^{-6}$. Now there is a more systematic progression, for example, for the ground state as one increases $G$; the 
wave function becomes more localized at  the centre of the square as $G$ increases. In fact this could be happening in a circular geometry, and the result would be the same.
As we move down, i.e. to more excited states for a given $G$, the wave function becomes more delocalized, and takes the shape of the edges in the square; these are veritable edge states. One notable exception is the result for $|\psi_{50}|^2$ with $G=50$, where the wave function becomes more concentrated; this is because we have suddenly moved on to the next Landau level (recall how we could track this explicitly in the case of the circular well). Note that in general the results compare quite well with those of the circular well [see Fig.~(\ref{prob_density1b})], except fo the obvious ``squareness'' of the contours of the higher excited states.}
\label{prob_density_sq2}
\end{figure}

\begin{figure}[H]
\begin{center}
\includegraphics[scale=0.45]{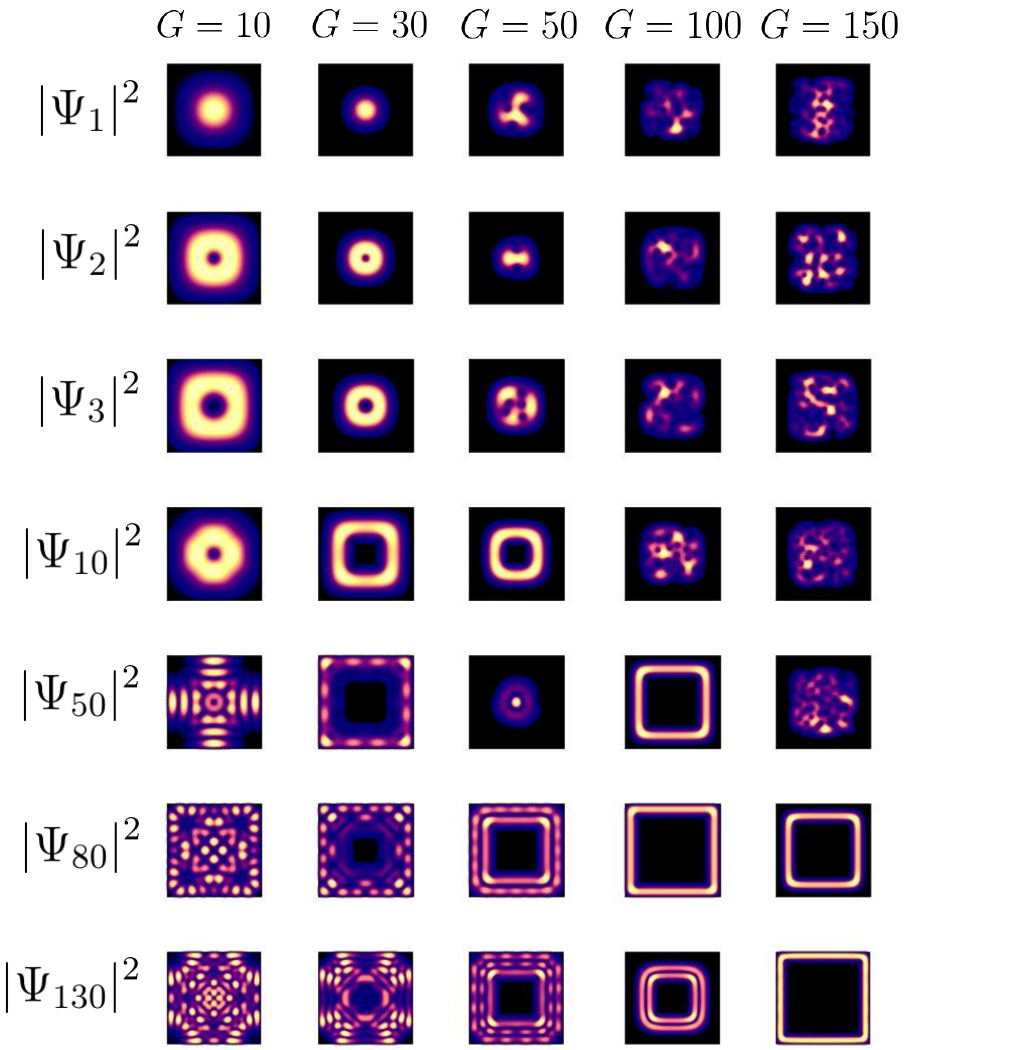}
\end{center}
\caption{Contour plots of the probability density, $|\Psi_{n}(x,y)|^2$, for various values of $n$, calculated with the Landau gauge, and as ordered by the diagonalization subroutine. The results agree with those in the symmetric gauge, Fig.~(\ref{prob_density_sq1}), only in the first two columns. Once again, the reason for the discrepancy in the remaining
columns is the degeneracy that is still present and this is remedied in the same manner as with the symmetric gauge results.}
\label{prob_density_sq_lan1}
\end{figure}

\begin{figure}[H]
\begin{center}
\includegraphics[scale=0.45]{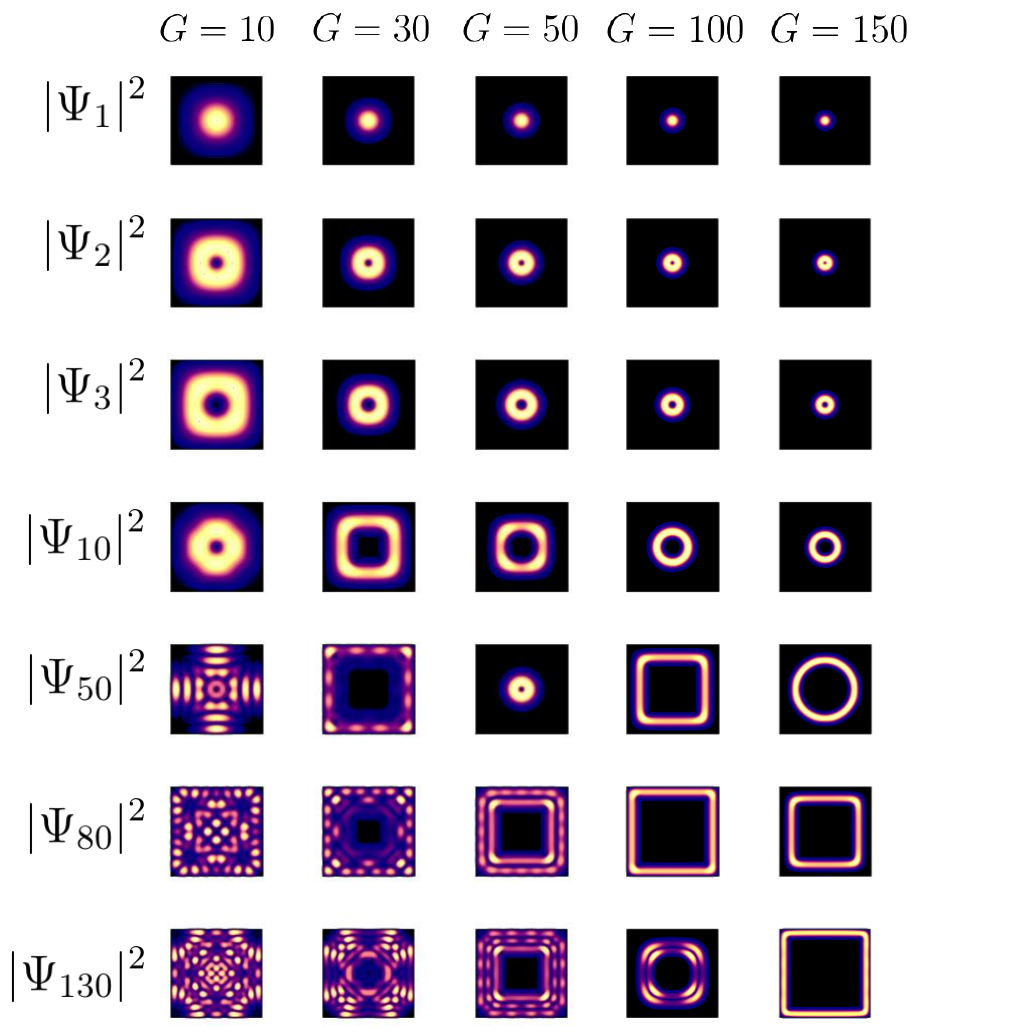}
\end{center}
\caption{Contour plots of the probability density in the Landau Gauge with the perturbative trap included. 
While these results are identical to those obtained in the symmetric gauge in Fig.~(\ref{prob_density_sq2}), we have
repeated them here to emphasize that this has been achieved with different gauges.}
\label{prob_density_sq_lan2}
\end{figure}

When this \textit{very weak} potential is included we obtain the results for the probability densities shown in Fig.~(\ref{prob_density_sq2}), where now the results look correct and more sensible
(the same perturbative potential was actually used in Fig.~(\ref{prob_density1b})). 
For the sake of completeness, we include the required matrix elements for our current problem,
\begin{equation}
\begin{split}
    {(h')}_{n,m}= \frac{H'_{n,m}}{E_0} = \frac{\pi^2 \epsilon^2}{4} &\Biggl[ \frac{1}{12} \delta_{n_x,m_x}\delta_{n_y,m_y} \left( 1 - \frac{6}{(\pi n_x)^2} + 1 - \frac{6}{(\pi n_y)^2} \right)\\[5pt]
    &+\frac{2}{\pi^2} \Bigl( \delta_{n_y,m_y} [1-\delta_{n_x,m_x}] g_e(n_x,m_x)  +  \delta_{n_x,m_x}  [1-\delta_{n_y,m_y}] g_e(n_y,m_y) \Bigr) \Biggr]
\end{split}
\label{ham_sq_pert}
\end{equation}
where $g_e(n,m)$ is as defined in Eq.~(\ref{ge_def}) and $\epsilon \equiv \hbar \omega_0/E_0 = (4G/\pi)(\omega_0/\omega_c)$; with this definition, $\epsilon^2$ provides 
the relative energy scale of the perturbing potential. These Hamiltonian matrix elements should be added to the previous ones given in Eq.~(\ref{ham_sq_dim}).\\
\\
\noindent\textbf{Landau Gauge}\\
Without using the perturbing potential in the Landau gauge, the probability density does not agree with that obtained in the Symmetric gauge. The results for the Landau gauge (without the perturbing potential) are shown in Fig.~(\ref{prob_density_sq_lan1}), where significant discrepancies with Fig.~(\ref{prob_density_sq1}) are apparent, particularly for larger values of $G$.
The results of Fig.~(\ref{prob_density_sq_lan1}) show somewhat random character, particularly for larger values of $G$, in a manner similar to that seen earlier [Fig.~(\ref{prob_density_sq1})] in the symmetric gauge, and for the same reason --- persistent numerical degeneracy. Nonetheless the addition of the perturbing potential, Eq.~(\ref{hamp}) with $\epsilon = 10^{-3}$ solves the problem as expected, and results identical (i.e. numerically, to about 5 digit accuracy) to those of Fig.~(\ref{prob_density_sq2}) are attained. These are shown in Fig.~(\ref{prob_density_sq_lan2}).

\subsection{\black{Probability Current in a Square Dot}}

Although probability current density is a gauge invariant quantity, the infinite degeneracy in free space results in two very different looking expressions for the Landau gauge and symmetric gauge. We have already resolved this issue for the probability densities in confined space, and so here we confirm that the current densities in each gauge are identical as well.

The expression for the probability current density is given by Eq.~(\ref{current1}). Adopting the expansion of the wave function in the square well (\ref{expand}) with basis states (\ref{basis_sq}) and arbitrary gauge choice $\mathbf{A}$, the probability current density of the $n$th total quantum state $\mathbf{J}_n$ can be expressed as:
\begin{align}
    \mathbf{J}_{n}(x,y)&= \frac{\hbar}{m_e}\left[{\rm Im}({\Psi}_n^{\ast} {\mathbf{\Theta}}_n) +\frac{e\mathbf{A}}{\hbar}|\Psi_n|^2, \right] \ \ \ \ \text{where}
    \label{sq_prob_curr_dens}\\[5pt]
    \Psi_n(x,y)&=\frac{2}{a} \sum_{m_x,m_y=1}^{\infty}c^{(n)}_{m_x,m_y} \sin\left(\frac{m_x\pi x}{a}\right)\sin\left(\frac{m_y\pi y}{a}\right), \ \ \ \ \text{and}\\[5pt]
    \begin{split}
     \mathbf{\Theta}_n(x,y)&\equiv \mathbf{\nabla}\Psi_n(x,y)=\frac{2}{a} \sum_{m_x,m_y=1}^{\infty}c^{(n)}_{m_x,m_y}\frac{\pi m_x}{a}\cos\left(\frac{m_x\pi x}{a}\right)\sin\left(\frac{m_y\pi y}{a}\right) \ \mathbf{\hat{x}}\\[5pt]
     &\ \ \ \ \ \ \ \ \ \ \ \ \ \ \ \ \ \ +\frac{2}{a}\sum_{m_x,m_y=1}^{\infty}c^{(n)}_{m_x,m_y}\frac{\pi m_y}{a}\sin\left(\frac{m_x\pi x}{a}\right)\cos\left(\frac{m_y\pi y}{a}\right) \ \mathbf{\hat{y}}.
     \end{split}
\end{align}
Note that the eigenvector coefficient corresponding to the $n^{\rm th}$ eigenvector, $c^{(n)}_{m_x,m_y}$, is in general a complex number. 

For the Symmetric [Eq.~(\ref{symm_shifted})] and Landau [Eq.~(\ref{landau_shifted})] gauge choices, the probability current density is given by
\begin{equation}
    \mathbf{J}^{(S)}_n(x,y)=\frac{\hbar}{m_e a}\left[a \ {\rm Im}({\Psi}_n^{\ast}{\mathbf{\Theta}}_n)+ \pi G|{\Psi}_n|^2\left(\left(\frac{x}{a}-\frac{1}{2}\right)\mathbf{\hat{y}}-\left(\frac{y}{a}-\frac{1}{2}\right)\mathbf{\hat{x}}\right)\right]
    \label{shifted_symm_prob_curr}
\end{equation}
for the shifted-symmetric gauge, and
\begin{equation}
    \mathbf{J}^{(L)}_n(x,y)=\frac{\hbar}{m_e a}\left[a \ {\rm Im}({\Psi}_n^{\ast}{\mathbf{\Theta}}_n)+ 2\pi G|{\Psi}_n|^2\left(\frac{x}{a}-\frac{1}{2}\right)\mathbf{\hat{y}} \right]
    \label{shifted_landau_prob_curr}
\end{equation}
for the shifted-Landau gauge. While the structure of the equations for the two gauge choices Eqs.~(\ref{shifted_symm_prob_curr}) and (\ref{shifted_landau_prob_curr}) 
is quite different, the probability current density is a gauge-invariant quantity, and we have confirmed that both results are identical.

\begin{figure}[H]
    \centering
    \includegraphics[scale=0.3]{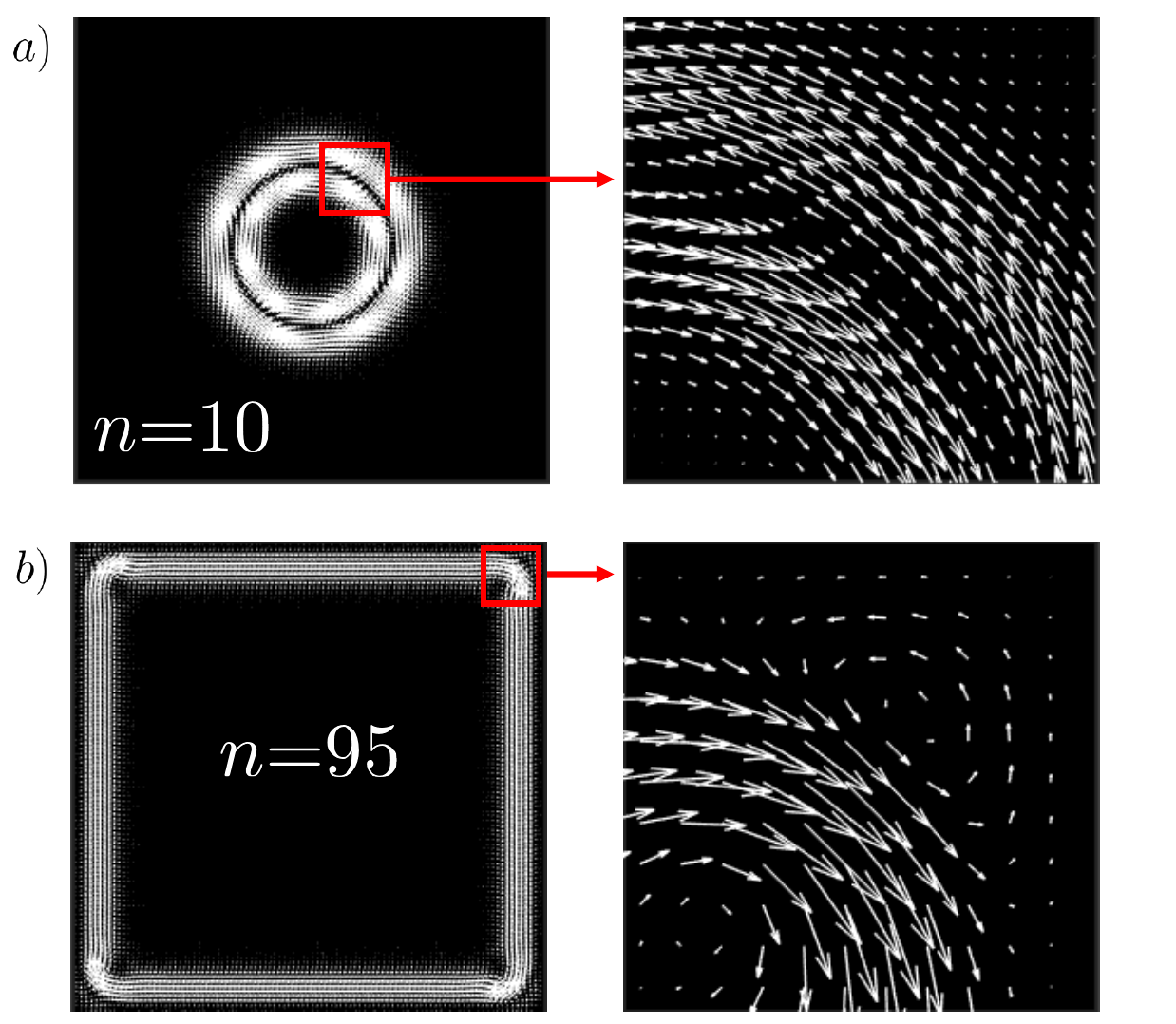}
    \caption{A vector field plot  of the probability current density with $G=100$ for a ``bulk'' state in (a) ($n = 10$) and an edge state in (b) ($n=95$).
    Further details of the region enclosed by the red square are provided in each case to the right. Convergence was attained with a matrix size of $N=150$ and $\epsilon=10^{-3}$ was used for the perturbing parabolic trap. Although this result was generated with the symmetric gauge, we confirmed that an identical result is 
    found for the Landau gauge. In the first case an inner shell of clockwise-circulating current is followed by a concentric shell of counter-clockwise-circulating current;
    this is very similar to what was found with circular geometry [see Fig.~(\ref{circ_vec_field}a)]. However, the edge state current density is very different. As in the circular
    case it is {\it primarily} a clockwise-circulating current, but the corners cause a vortex-anti-vortex pair to be created, as is clear from the blow-up on the right.}
    \label{fig: square_well_vec_field}
\end{figure}

Equipped with equations~(\ref{sq_prob_curr_dens}-\ref{shifted_landau_prob_curr}) it is
straightforward to evaluate the probability current density for a given eigenstate. Two vector field plots of typical results are shown in Fig.~(\ref{fig: square_well_vec_field})
(in units of $h/(m_e a^3)$).
We see that (a)
is a ``bulk'' state while (b) is an edge state. In fact, the former result is qualitatively indistinguishable from that attained for a circular 
geometry in Fig.~(\ref{circ_vec_field}), as is apparent from
the result. In all cases we have also used a perturbing potential as described earlier, with $\epsilon = 10^{-3}$. In (b) there is a clear difference from the circular
case, and the current tends to follow the geometry of the square boundary. While one may suspect from this that edge state currents simply follow the geometry of the boundary, a closer look at the expanded portion in (b) shows very interesting structure. Rather than smoothly following the geometry of the boundary at the corners of the square well, the probability current density appears to form stationary vortices rotating in a direction opposite to that of the nearby principal current density flow. This is a very different
scenario than that encountered with the circular geometry, and is reminiscent of the `corner modes'  noted recently for topological insulators.\cite{yan18} Here, these are not
zero energy modes, but appear as a simple consequence of the square geometry, and occur at large quantum number. Thus they may in fact be the precursor of
semi-classical behaviour (e.g. the skipping orbits described in Ref.~[\onlinecite{lent91}]). Further investigation is clearly required. 

\begin{figure}[H]
    \centering
    \includegraphics[scale=0.25]{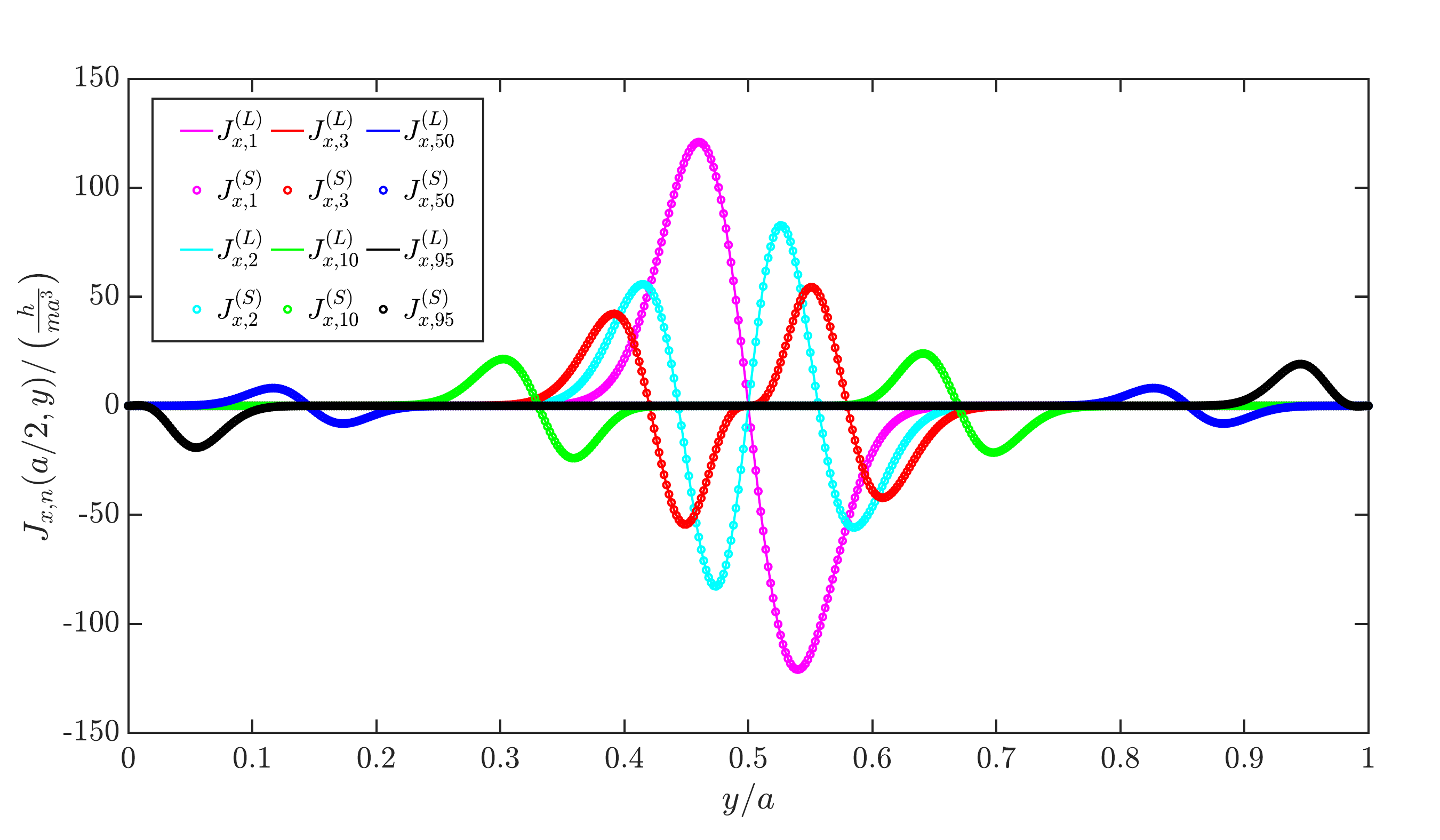}
    \caption{The $x$-component of the current density $J_{x,n}$ as a function of $y$ taken through $x=a/2$. This represents a vertical slice through the centre
    of the sample. We show results for various eigenvalue numbers $n$, as labeled. Results obtained in the Landau (Symmetric) gauge are shown as curves (points),
    and obviously agree with one another. Here we used $G=100$ and $N=150$. Also note that $J_{y,n}\sim0$ over this slice.}
    \label{fig: x=a/2_prob_curr_slice}
\end{figure}

One can also characterize the probability current density at slices through the sample.
Defining $J_{x,n}\equiv\mathbf{J}_n\cdot\mathbf{\hat{x}}$, and $J_{y,n}\equiv\mathbf{J}_n\cdot\mathbf{\hat{y}}$,
we show $J_{x,n}$ as a function of $y$ (a vertical slice at $x=a/2$) in Fig.~(\ref{fig: x=a/2_prob_curr_slice}) through the centre of the square. For low quantum numbers the result will resemble that of the circle, shown in Fig.~(\ref{prob_current2}). Here we cannot
classify the states according to their $\ell$ quantum number, as we did in that case. Also note that Fig.~(\ref{prob_current2}) displays results across a radius, i.e.
{\it half} the sample, whereas for the square, Fig.~(\ref{fig: x=a/2_prob_curr_slice}) shows results across {\it the entire} sample, and therefore has an inherent asymmetry, This is
because, in spite of the {\it square} geometry, the current is {\it circulating}. For the same reason, $J_{y,n}$ along this vertical slice was found to be zero (within
numerical noise). Slices related through symmetry operations yield entirely equivalent results. For example, the $J_{y,n}$ plotted
across a horizontal slice at $y=a/2$ looks identical to $J_{x,n}$ at $x=a/2$.
As seen before in the circular well, the current contribution for a particular state becomes non-zero as we move from the bulk region to the edge region
in Fig.~(\ref{fig: x=a/2_prob_curr_slice}). Moving from $y=a/2$ to $y=a$, we see that the total current over this region vanishes for the bulk states (with the exception of the ground state), due to a sign change across a node ($J_{x,10}$, for example, has equal amounts of positive and negative current density between $y=a/2$ and $y=a$). States near the edge ($J_{x,95}$, for example) do not experience this nodal sign change, resulting in a non-zero current contribution, as indicated by a positive-only current density near $y=a$. Also note that the edge state current has opposite sign on opposite edges of the confinement, confirming its inherent chirality.\\
\begin{figure}[H]
    \centering
    \includegraphics[scale=0.25]{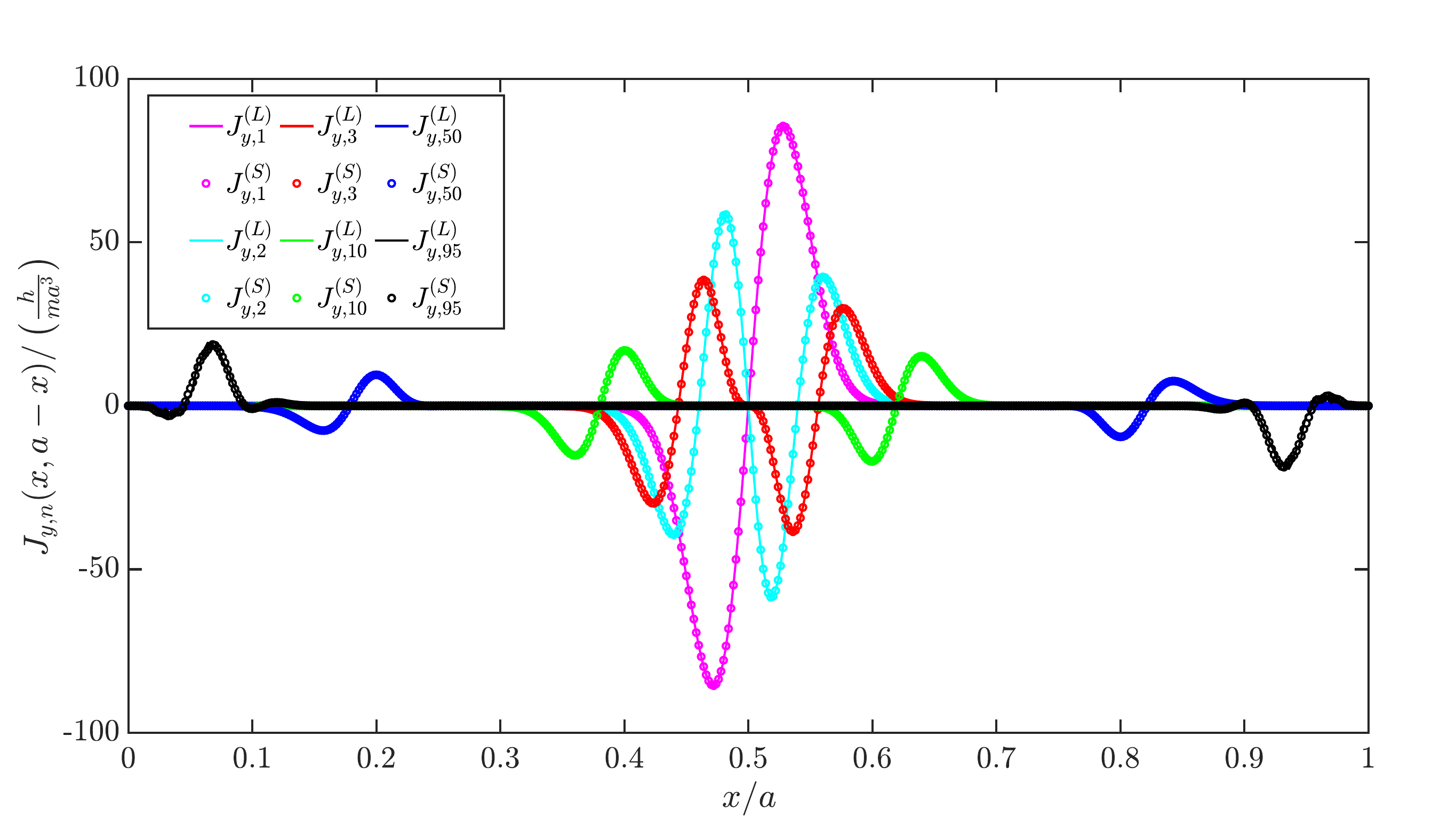}
    \caption{$J_{y,n}$ as a function of $x$, taken along the diagonal
    from the upper left of the square down to the lower right. As $x$ varies from $0$ to $a$, $y = a-x$ varies also, from $a$ to $0$. As in Fig.~(\ref{fig: x=a/2_prob_curr_slice}),
    results are shown for various quantum numbers and in both gauges. We used $G=100$ and $N=150$. $J_{x,n}$ mirrors this result about the vertical axis. }
    \label{fig: diagonal_prob_curr_slice}
\end{figure}
In Fig.~(\ref{fig: diagonal_prob_curr_slice}) we show results across another slice, a diagonal across the square from the upper left to the lower right. We show only $J_{y,n}$ but in this case $J_{x,n}$ is simply the mirror of $J_{y,n}$ about the vertical axis. This plot is not so different from the previous one, but can
give us a glimpse of the vorticies seen only at the corners of the square. The occurrence of these vortices clearly merits further study.

\subsection{\black{Other Gauge Possibilities}}

Finally, we should note that we explored other gauge choices, namely
 \begin{equation}
   \mbf{A}^{\prime\prime}_{S}=\frac{B}{2}\left[\left(x-\frac{b_x}{2}\right)\hat{\mbf{y}}-\left(y-\frac{b_y}{2}\right)\hat{\mbf{x}}\right],
   \label{symm_shiftedb}
\end{equation}
for the symmetric gauge, and
 \begin{equation}
   \mbf{A}^{\prime \prime}_L=B\left(x-\frac{b_x}{2}\right)\hat{\mbf{y}}
   \label{landau_shiftedb}
\end{equation}
for the Landau gauge, where $b_x$ and $b_y$ are {\it arbitrary}. In particular we formulated the problem for $b_x = b_y = 0$, knowing that this was
putting ourselves at a disadvantage. Nonetheless, even this choice works, but at a cost that much larger matrices are required. This is expected, since now the
gauge potential is very asymmetric throughout the square --- in either case the parabolic confinement arising from the applied magnetic field
is centred at one corner of the square. While we refrain from showing results here, we found identical results as earlier for weak magnetic fields (where
we could pursue convergence as a function of matrix size). It is noteworthy that in this regime there is no practical degeneracy and so the perturbation potential,
Eq.~(\ref{hamp}), is not required.

\section{\black{Summary}}

We have presented an elementary discussion of the issues concerning gauge invariance, first in free space, and, more pertinently, in confined geometries.
Practical degeneracies still exist in confined geometries, or quantum dots, in particular where a level ``floor'' in the potential well exists; this is not the case
with a parabolic trap, for example. We used a quantum dot with circular symmetry to illustrate some of these degeneracies. However, in this case it is
easy to overlook some of the subtleties, as we conveniently have a good quantum number, $\ell$, that can both simplify the calculation, and can be used to organize
the results in an unambiguous manner. In this way, even states whose energy difference cannot be distinguished to numerical precision will nonetheless
be ordered properly through our knowledge of these quantum numbers.

The circular potential also ``begs'' for the Symmetric Gauge to be utilized. Doing so simplifies the problem immensely, so only a one-dimensional
equation requires solution. We could have used the Landau gauge, but then the problem would have been significantly more difficult. For the case of
a two-dimensional square, however, the problem is more difficult right from the start, regardless of which gauge is chosen. Partly for this reason, it became a good
testing ground for comparing gauge choices, especially given our method of solution, matrix mechanics. To our knowledge this problem has not been previously solved
in either gauge.

Choosing a geometry also highlights that a gauge choice also involves a choice of origin for the gauge. Whether we use the Symmetric or Landau gauge, 
the natural origin is the centre of the sample, but we showed that this is not required. Non-optimal choices (i.e. {\it not} the centre of the sample) generally
require more basis states with our method, so there is an additional numerical cost for a non-optimal choice. On the flip side, our method still allows calculation in
{\it any} gauge, so students can readily check various gauge choices. One can compute any property desired (we showed probability and
current densities in this study), since we have the eigenvalues and eigenvectors. A further study will explore the susceptibility and other properties that
can readily be measured.

Finally, we also suggested a simple way of removing the degeneracy at the $10^{-10}$ level (so the computer could tell the difference). As plotted, the eigenvalues
will still appear to be degenerate, and this is the correct physics. However, for purposes of organization it is necessary to have a method to distinguish these from one another,
and the (very) shallow gauge-invariant parabolic potential that we proposed does the trick.

The results for a square geometry are new; we expected that, for the properties studied in this work, the square geometry would not produce anything qualitatively new 
beyond the results for the circular quantum dot. Instead, as we saw, the current density for the edge states contains vortex-anti-vortex pairs at the corners, which are completely absent for the circular dot for the same range of quantum numbers. We plan to carry out a more in-depth investigation of the conditions under which such modes appear.

\section{\black{Acknowledgments}}

This work was supported in part by the Natural Sciences and Engineering
Research Council of Canada (NSERC). These studies were initiated through a number of former students; in particular
Sophie Taylor carried out some initial studies that were very helpful to get us started. We also thank Ania Michalik for helpful
calculations and discussions. Mason Protter and Joel Hutchinson were instrumental in the initial stages with assistance with numerically technical matters.
We also thank Joseph Maciejko and Jorge Hirsch for very helpful discussions on various aspects of this problem.

\appendix

\section{\black{Review of Landau Levels in Free Space}}

\subsection{\black{The Symmetric Gauge}}
\noindent\textbf{Eigenstates}\\
Substituting $\mathbf{A}_S$ into the Hamiltonian (\ref{ham_original}) leads to
\begin{equation}
H=\frac{\mathbf{p}^2}{2m}-i\frac{\hbar\omega_c}{2}(x\partial_y-y\partial_x)+\frac{1}{8}m\omega_c^2(x^2+y^2),
\label{ham_symm1}
\end{equation}
where $\partial_x \equiv \partial/\partial x$, and similarly for $y$ and later for radial coordinates as well. We have introduced
the classical cyclotron frequency $\omega_c \equiv {eB}/{m}$. Converting to polar coordinates, we get
\begin{equation}
    H=-\frac{\hbar^2}{2m}(\partial_r^2+\frac{1}{r}\partial_r)+\frac{L_z^2}{2mr^2}+\frac{\omega_c}{2}L_z+\frac{1}{8}m\omega_c^2r^2,
\label{ham_symm2}
\end{equation}
where $L_z= -i\hbar (x\partial_y - y \partial_x) = -i\hbar \partial_{\phi}$ is the standard operator for the $z$ component of the angular momentum. Note that this operator is \textit{gauge-dependent} and its eigenvalues should not be considered to represent the real, physical angular momentum of the electrons in this system. To consider the \textit{true} angular momentum of the system, we must take into account the effect of the magnetic field. Just like the canonical momentum is shifted to the kinetic momentum in the presence of a magnetic field, we can also define a kinetic angular momentum $\mathbf{\Lambda}$,
\begin{equation}
    \mathbf{L}=\mathbf{r}\cross\mathbf{p}\rightarrow\mathbf{\Lambda}=\mathbf{r}\cross\boldsymbol{\pi}=\mathbf{r}\cross(\mathbf{p}+e\mathbf{A})=\mathbf{L}+\mathbf{r}\cross e\mathbf{A}.
   \label{kin_ang_mom}
\end{equation}
Since we are working in 2 dimensions, the kinetic angular momentum is perpendicular to the plane and is given by
\begin{equation}
    \mathbf{\Lambda}(\mathbf{r})=\left(L_{z}+\frac{1}{2}m\omega_cr^2\right)\boldsymbol{\hat{z}}=\Lambda_z(r)\boldsymbol{\hat{z}},
    \label{kin_ang_mom2}
\end{equation}
where $\Lambda_z(r)$ is the magnitude of the kinetic angular momentum in the $\boldsymbol{\hat{z}}$ direction.
We can also rewrite Eq.~~(\ref{ham_symm2}) in terms of $\Lambda_z(r)$,
\begin{equation}
    H=-\frac{\hbar^2}{2m}(\partial_r^2+\frac{1}{r}\partial_r)+\frac{\Lambda_z^2(r)}{2mr^2}.
       \label{ham_symm3}
\end{equation}
Analogously, this new term can be interpreted as a kinetic centrifugal potential $V_{l}(r)$.

Returning to our initial problem, to calculate the eigenstates and eigenvalues of Eq.~~(\ref{ham_symm2}), we first note that since $[H,L_z]=0$, we can write the eigenstates $\Psi(r,\phi)$ as
\begin{equation}
    \Psi(r,\phi)=e^{i \ell \phi}\psi(r),
    \label{separable}
\end{equation}
where $\hbar \ell$ is the eigenvalue of the $L_z$ operator. 
Using periodic boundary conditions in the variable $\phi$, $\Psi(r,\phi+2\pi)=\Psi(r,\phi)$, requires
$\ell$ to be an integer, and the eigenvalues of $L_z$ are thus quantized, with $\ell$ the `azimuthal' or `angular momentum' quantum number. Inserting
Eq.~(\ref{separable}) into Eq.~(\ref{ham_symm2}) results in the radial ordinary differential equation,
\begin{equation}
      \left[-\frac{\hbar^2}{2m}\left(\frac{d^2}{dr^2}+\frac{1}{r}\frac{d}{dr}\right)+\frac{\hbar^2 \ell^2}{2mr^2}+\frac{1}{2}m\left(\frac{\omega_c}{2}\right)^2r^2 +\frac{\hbar \omega_c\ell}{2} \right]\psi(r) = E\psi(r).
      \label{ham_rad}
\end{equation}
Eq.~(\ref{ham_rad}) is precisely of the form of a 2D isotropic radial quantum harmonic oscillator (QHO) Hamiltonian with frequency $\omega_c/2$, with an energy shift of ${\hbar\omega_c\ell/2}$.
Although the solution to the above is known well in the literature, often only operator algebra-based derivations are shown (see Refs.~[\onlinecite{jain2007composite}, \onlinecite{qheTong}, \onlinecite{Murayama}] for example). While those derivations avoid brute force series methods, the behaviour of the eigenstate is not so clear from the resulting operator-based expressions. For this reason, we will carry out an explicit derivation of the eigenstate here.

We first introduce a characteristic magnetic length scale,  $\ell_B\equiv\sqrt{{\hbar/(eB)}}$, and make a change of variables:
\begin{equation}
x = \left(\sqrt{\frac{m\omega_c}{2\hbar}}\right)r \equiv r/(\sqrt{2}\ell_B).
\label{r to x}
\end{equation}
The length $\ell_B$ represents a characteristic length scale for the wave function of a charged particle in a magnetic field.  For example, for $B\approx1$ Tesla , $l_B\approx26$ nm.\\ 
This change-of-variables simplifies Eq.~ (\ref{ham_rad}) to the following:
\begin{equation}
    \left[\frac{d^2}{dx^2}+\frac{1}{x}\frac{d}{dx}-\frac{\ell^2}{x^2}-x^2+K_\ell\right]\psi(x) = 0
    \label{symm_der1}
\end{equation}
where $K_\ell \equiv 2(2E/(\hbar\omega_c)-\ell)$.
We simplify further with the following functional substitution:
\begin{equation}
    \psi(x) = u(x)/\sqrt{x}
\end{equation}
converting our original ordinary differential equation (ODE) in Eq.~ (\ref{symm_der1}) to a 1D-like radial Schr\"{o}dinger equation:
\begin{equation}
    \left[\frac{d^2}{dx^2} - \frac{\ell^2-1/4}{x^2}-x^2+K_\ell\right]u(x) = 0.
    \label{symm_der2}
\end{equation}
In the above equation, one can identify (convergent) asymptotic solutions for large and small $x$:
\begin{align}
    \text{For} \ x\to\infty,& \ u''(x)\sim x^2u(x) \Rightarrow u(x) \approx e^{-\frac{x^2}{2}} \\  
    \text{For} \ x\to 0,& \ u''(x)\sim \frac{\ell^2-1/4}{x^2} u(x)\Rightarrow u(x) \approx x^{|\ell|+1/2}.
\end{align}
The above equations motivate the following ansatz for $u(x)$:
\begin{equation}
    u(x) = e^{-\frac{x^2}{2}}x^{|\ell|+1/2}v(x)
    \label{u to v}
\end{equation}
Substituting the above into Eq.~ (\ref{symm_der2}) leads to:
\begin{equation}
    v''(x)+2\left[\frac{|l|+1/2}{x}-x\right]v'(x)+\left[K_\ell-2(|l|+1)\right]v(x) = 0.
    \label{symm_der3}
\end{equation}
With another change of variables $x = \sqrt{y}$, the above equation reduces to a not-so-familiar ODE:
\begin{equation}
    yv''(y)+\left[|\ell|+1-y\right]v'(y)+(1/4)[K_\ell-2(|\ell|+1)]v(y) = 0.
    \label{symm_der4}
\end{equation}
This ODE has the form of a confluent hypergeometric equation, which can be written as (see Ref.~\onlinecite{NIST:DLMF}, section 13.2):
\begin{equation}
    zw''(z)+(b-z)w'(z)-aw(z) = 0.
    \label{CHG_eq}
\end{equation}
There are two independent solutions to this equation; one is discarded because of the required behavior at the origin, and the other is the so-called Kummer function:
\begin{equation}
    w(z) = M(a,b;z) =\sum_{j=0}^{\infty} \frac{(a)_j z^{j}}{(b)_j j !},
    \label{CHG_func}
\end{equation}
where the notation $(a)_j$ is the so-called
Pochhammer's symbol, which can be computed as follows:
\begin{equation}
\begin{split}
    (a)_j &= a(a+1)(a+2) \cdots(a+j-1), \ \ j=1,2,3,...\\[5pt]
     (a)_0 &= 1.
    \label{rising_fact}
\end{split}
\end{equation}
Comparing Eqs. (\ref{symm_der4}) and (\ref{CHG_eq}), it is clear that we may write the solution to Eq.~ (\ref{symm_der4}) as:
\begin{equation}
    v(y) = M\left(-\frac{K_\ell-2(|\ell|+1)}{4},|\ell|+1; y\right).
\end{equation}
As seen in Eq.~(\ref{CHG_func}), our solution for $v(y)$ is an infinite series. This leads to a troublesome issue: for large argument $y$, $v(y)$ will have asymptotic behaviour like:
\begin{equation}
v(y) = \sum_{j=0}^{\infty} \frac{[-(K_\ell-2(|\ell|+1)/4]^{(j)}}{[|\ell|+1]^{(j)}  }\frac{y^{j}}{j!} \sim \sum_{j=0}^\infty \frac{y^j}{j!} = e^y\to\infty, \ \text{for} \ y\to\infty
\end{equation}
Thus, we see that $v(x)\sim e^{x^2}$ after returning to the $x$ variable via $y=x^2$. This implies that our solution for $u(x)$ will have divergent asymptotic behaviour for large argument:
\begin{equation}
    u(x) = e^{-x^2/2}x^{|\ell|+1/2}v(x) \sim e^{-x^2/2}x^{|\ell|+1/2}e^{x^2} = x^{|\ell|+1/2}e^{x^2/2}\to\infty, \ \text{for $x\to\infty$},
\end{equation}
and in particular, the squared magnitude of this function is not integrable.
This problem can be remedied in the usual way, by requiring that the first parameter of Kummer's function ($a$ in Eq.~\ref{CHG_func}) is a non-positive integer $-n_r$, where $n_r=0,1,2...$. This condition truncates the infinite series since, by Eq.~ (\ref{rising_fact}):
\begin{equation}
       (-n_r)_{j} = 0 \ \text{for} \ j>n_r, \ n_r=0,1,2,....
\end{equation}
Thus, we demand that the following condition must hold for $u(x)$ to be normalizeable:
\begin{equation}
    (1/4)[K_\ell-2(|\ell|+1)] = n_r, \ n_r=0,1,2,...
    \label{quantization}
\end{equation}
As $K_\ell$ is defined in terms of $E$, this condition quantizes our allowed energies.
We see that after 
enforcing Eq.~ (\ref{quantization}), Eq.~ (\ref{symm_der4}) reduces to the generalized Laguerre equation:\cite{Weisstein}
\begin{equation}
    yv''(y)+\left[|\ell|+1-y\right]v'(y)+n_rv(y) = 0.
    \label{Laguerre_equation}
\end{equation}
The solution to this equation is given by:
\begin{equation}
    v(y) = A_{n_r,\ell}L_{n_r}^{|\ell|}\left[y\right],
\end{equation}
where $L^{\alpha}_n[z]$ is the generalized Laguerre polynomial and $A_{n_r,\ell}$ is a constant to be determined by normalization. The generalized Laguerre polynomials\footnote{Note that there are multiple definitions used for Laguerre polynomials in literature. The Laguerre polynomials are usually defined in physics texts in a manner that often differs from alternate definitions by a factorial factor. For example, $^{\mathrm{Arf}} {L_{n-\ell-1}^{2 \ell+1}}=$ $\frac{1}{(n+\ell) !} ^{\mathrm{Arf}} L_{n-\ell-1}^{2 \ell+1},$ so, in our case, $^{\mathrm{Arf}} L_{n-1}^{1}=\frac{1}{n !} ^{\mathrm{Griff}} L_{n-1}^{1}$
where $^{\mathrm{Griff}} L_{n-1}^{1}$ denotes the notation used by Griffiths\cite{griffiths18} and $^{\mathrm{Arf}} L_{n-1}^{1}$ denotes the notation used by Arfken, in G. Arfken, Mathematical Methods for Physicists, Third Edition, (Academic Press, Toronto, 1985 ). In Eq.~(\ref{Laguerre_norm}), we use the definition from Arfken.} obey the following orthogonality condition over $z\in[0,\infty)$:\cite{Weisstein}
\begin{equation}
    \int_0^\infty e^{-x}x^{\alpha}L_{n'}^{\alpha}[z]L_{n}^{\alpha}[z]dz = \frac{(n+\alpha)!}{n!}\delta_{n'n}, \ \ \ \alpha>0. 
    \label{Laguerre_norm}
\end{equation}
Changing variables back to $x$ from $y$, we acquire the solution to Eq.~ (\ref{symm_der2}) through Eq.~ (\ref{u to v}):
\begin{equation}
    u(x) = A_{n_r,\ell} e^{-\frac{x^2}{2}}x^{|\ell|+1/2}L_{n_r}^{|\ell|}\left[x^2\right].
\end{equation}
With the above, we obtain the  solution to our original radial ODE (Eq.~ \ref{ham_rad}) by undoing our functional substitution $\psi(x) = u(x)/\sqrt{x}$ and changing variables back to $r$ from $x$ using Eq.~ (\ref{r to x}): 
\begin{equation}
    \psi_{n_r,\ell}(r)=A_{n_r,\ell}\left(\frac{r}{\sqrt{2}l_B}\right)^{|\ell|}e^{-\frac{r^2}{4l_B^2}}L^{|\ell|}_{n_r}\left[\frac{r^2}{2l_B^2}\right].
    \label{rad_soln}
\end{equation}
Using Eq.~ (\ref{Laguerre_norm}), the normalization constant $A_{n_r,\ell}$ can be found:
\begin{equation}
    \begin{split}
        1 &= A_{n_r,\ell}^2\int_0^\infty |\psi_{n_r,\ell}(r)|^2rdr = A_{n_r,\ell}^2\int_0^\infty   \left(\frac{r^2}{2l_B^2}\right)^{|\ell|}e^{-\frac{r^2}{2l_B^2}}\left(L^{|\ell|}_{n_r}\left[\frac{r^2}{2l_B^2}\right]\right)^2rdr\\[5pt]
        &= A_{n_r,\ell}^2l_B^2\int_0^\infty x^{|\ell|}e^{-x}\left(L^{|\ell|}_{n_r}\left[x\right]\right)^2dx = A_{n_r,\ell}^2l_B^2\frac{(n_r+|\ell|)!}{n_r!}\\[5pt]
        \Rightarrow A_{n_r,\ell} &= \frac{1}{l_B}\sqrt{\frac{n_r!}{\left(n_r+|\ell|\right)!}}
    \end{split}
\end{equation}
Reattaching the angular component $e^{i\ell\phi}$ and normalizing over $\phi\in[0,2\pi]$, we finally obtain the normalized eigenstate for the symmetric gauge Hamiltonian (Eq.~ \ref{ham_symm2}):

\begin{equation} 
    \Psi_{n_r,\ell}(r,\phi)=\frac{1}{l_B}\sqrt{\frac{n_r!}{\left(n_r+|\ell|\right)!}}\left(\frac{r}{\sqrt{2}l_B}\right)^{|\ell|}e^{-\frac{r^2}{4l_B^2}}L^{|\ell|}_{n_r}\left[\frac{r^2}{2l_B^2}\right]\frac{e^{i\ell\phi}}{\sqrt{2\pi}}.
    \label{psi_rad}
\end{equation}
The quantum number $n_r$ is the `radial quantum number' and simply counts the number of nodes that the radial part of the wave function has ($n_r$ is a non-negative integer). The corresponding probability density takes the appearance of concentric rings that are radially localized increasingly outwards from $r=0$ with increasing $|\ell|$. The wave
function has $n_r$ nodes and the probability density $|\Psi|^2$ depends on $|\ell|$ and \textit{not the sign} of $\ell$. Examining the wave function alone, its dependence on the value of $\ell$ may lead one to think that there is a double degeneracy for each energy. This is not the case here, as explained in the following subsection.\\
\newline
\noindent\textbf{Eigenenergies}\\
The expression for the eigenenergies corresponding to the eigenstates given by Eq.~(\ref{psi_rad}) are obtained by rearranging Eq.~(\ref{quantization}) for $E$:
\begin{equation}
    E_{n_r,\ell} = \frac{\hbar\omega_c}{2}(2n_r+|\ell|+1) + \frac{\hbar \omega_c\ell}{2},
    \label{eig1}
\end{equation}
where the appropriate quantum numbers are now subscripted.
In Eq.~(\ref{eig1}), the first term gives the standard eigenenergy of a 2D radial QHO (albeit with frequency $\omega_c/2$). The second term comes from the coupled momentum-gauge term that manifests in the Hamiltonian after expanding the kinetic momentum operator. This term represents the interaction between the canonical momenta and magnetic field, and is what distinguishes this problem from a standard 2D isotropic QHO. 
If we combine these two terms, we get:
\begin{equation}
     E_{n_r,\ell} = \hbar\omega_c\left(n_r+\frac{|\ell|+\ell}{2}+\frac{1}{2}\right)
\end{equation}
so for all $\ell \leq 0 \ $, $E_{n_r,\ell} = \hbar\omega_c\left(n_r+\frac{1}{2}\right)$, and the energy spectrum is in fact infinitely degenerate.
This infinite degeneracy is another distinct feature of the problem of a free electron in a magnetic field.

A new quantum number 
$n_L \equiv n_r+{(|l|+l)}/{2}$ can be defined as well, and then the energies are simply
\begin{equation}
    E_{n_L}=\hbar\omega_c\left(n_L+\frac{1}{2}\right)
    \label{energy_landau_level}
\end{equation}
where $n_L=0,1,2,...$ is the Landau level quantum number. 
Clearly, further degeneracies are possible through the use of $\ell > 0$ for $\ell \leq n_L$. This is shown in the table below, where all
energies have an infinite degeneracy due to the negative $\ell$ states that carry on indefinitely to the right for each energy. Additionally,
however, as $n_L$ increases, positive $\ell$ states are also degenerate in growing number as indicated.
\begin{equation}
\begin{aligned}
{\rm Energy} \ (\hbar \omega_c/2) \phantom{aaaaaaaa} n_L \phantom{aaaaaaaaaaaaaaaaaaaaa} (n_r,\ell) \ \ \phantom{aaaaaaaaaaaaaa} \cr
\vdots \phantom{aaaaaaaaaaaaaaaa} \vdots \phantom{aaaaaaaaaaaaa} \phantom{(0,2) \ \ (1,1) \ \ } \vdots \phantom{\ \ (2,-1) \ \ (2,-2)} \dots \cr
5        \phantom{aaaaaaaaaaaaaaaa} 2         \phantom{aaaaaaaaaa} (0,2) \ \ (1,1) \ \  (2,0) \ \ (2,-1) \ \ (2,-2) \dots \cr
3 \phantom{aaaaaaaaaaaaaaaa} 1        \phantom{aaaaaaaaaa} \phantom{(0,2)} \ \ (0,1) \ \  (1,0) \ \ (1,-1) \ \ (1,-2) \dots \cr 
1 \phantom{aaaaaaaaaaaaaaaa} 0 \phantom{aaaaaaaaaa} \phantom{(0,2) \ \ (1,1)} \ \  (0,0) \ \ (0,-1) \ \ (0,-2) \dots \cr
\end{aligned}
\end{equation}\\
\\
This infinite degeneracy exhibits a peculiar pattern that is asymmetric with the sign of $\ell$. The electron, in response to the magnetic field, appears to 'prefer' a direction for $L_z$, requiring more energy to exist in the positive $\ell$ states than the negative $\ell$ states. Physically, circulating with positive angular momentum in the presence of a magnetic field directed in the positive $z$ direction costs energy for an electron, whereas circulating with negative angular momentum does not. This is consistent with the classical idea that electrons speed up or slow down depending on whether they are circulating with positive or negative angular momentum, respectively.\cite{griffiths99}\\
\newline
\noindent\textbf{The Lowest Landau Level}\\
To better understand the wave functions and issues associated with them we focus on the lowest Landau level (LLL). For the LLL, $n_r = 0$ and $\ell \leq0$. Then $E_0 = \hbar \omega_c/2$ and the associated Laguerre polynomial is identically unity. So an infinite set of degenerate LLL wave functions is given simply by
\begin{equation}
    \Psi_{LLL}(r,\phi)=\frac{e^{i\ell\phi}}{\sqrt{2\pi}}\left(\frac{r}{\sqrt{2}l_B}\right)^{|\ell|}\frac{e^{-\frac{r^2}{4l_B^2}}}{l_B\sqrt{|\ell|!}},  \ \ \ \ \ \ell \le 0.
    \label{lll_wavefunction}
\end{equation}
The corresponding probability density $|\Psi_{LLL}|^2$ is a radial Gaussian function centred around $r=0$ but with a maximum at 
$r_{\ell}=l_B\sqrt{2|\ell|}$ with a spread about this maximum of $\sim\mathcal{O}(l_B)$. As $|\ell|$ increases, the location of the maximum of $|\Psi_{LLL}|^2$, i.e. $r_{\ell}$, moves outwards {\it from zero}. On the other hand, increasing the magnetic field $B$ (thus decreasing $l_B$) leads to a decrease in the position of the maximum in $|\Psi_{LLL}|^2$. Also note that as $|\ell|$ increases the difference between centres of two successive $\ell$ eigenstates $\Psi_{n_r,\ell}$ and $\Psi_{n_r,\ell+1}$
will be much less than the spread $l_B$ [see Eq.~~(\ref{lll_wavefunction})]  and will decrease like 
$1/\sqrt{|\ell|}$ for large $|\ell|$, implying that the corresponding eigenstates will increasingly overlap with one another.
The values of $r_{\ell}$ also reveal that the amount of magnetic flux
$\Phi_B$ associated with each wave function in the LLL is quantized in units of the magnetic flux quantum $\Phi_0={h}/{e}$. This is easily seen: consider the magnetic 
flux $\Phi_B$ between a circle of radius $r_{\ell}$ and that of (smaller) radius $r_{\ell + 1}$ (recall that $\ell \leq 0$, so that $r_{\ell} > r_{\ell +1}$ ):
\begin{equation}
    \Phi_B=B\pi (r_{\ell}^2-r_{\ell+1}^2) = 2 \pi B l_B^2(|\ell| - |\ell+1|) = 2 \pi B\frac{\hbar}{eB}=\Phi_0.
    \label{flux_quantum}
\end{equation}
Thus, there is exactly a single quantum of magnetic flux between successive (negative) $\ell$ states in the LLL.
For the quantum dot geometries considered in sections III and IV in this paper, the system is finite, and the electron is confined to some extent, so there must be some finite number of $\ell$'s that constitute the degeneracy number for a given Landau level (since the wave functions grow outward with increasing $|\ell |$). The standard argument (already provided above) is that the expected degeneracy $G$ of the $LLL$, given a sample of area $A$, is the ratio of the total magnetic flux $\Phi$ to the flux quantum,
\begin{equation}
    G=\frac{\Phi}{\Phi_0}=\frac{BA}{h/e}.
    \label{degeneracy}
\end{equation}
As demonstrated later, this is not quite true, as sample edges confine the electron and result in higher energies for states near the edges.\\
\newline
\noindent\textbf{Gauge Invariance and Choice of Origin}\\
It is important to note that these results are all gauge-dependent, even given that we use the Symmetric gauge. In particular, the choice of Symmetric gauge with $A_s$ defined in the first paragraph of the introduction represents just one possible choice (out of infinitely many) where the vector field is referred to the origin. Since for this problem there is no preferred origin the wave functions will reflect this choice and be gauge-dependent. For this reason we highlighted ``{\it from zero}'' in italics above. An alternative would have been to use
\begin{equation}
    \mathbf{A}_S=\frac{B}{2}[(x-x_0)\boldsymbol{\hat{y}}-(y-y_0)\boldsymbol{\hat{x}}],
        \label{gauge_as}
\end{equation}
and then the wave functions would have been different, even though this represents the same problem, i.e. a charged particle in free space in the presence of a uniform magnetic field $B$ (see, for example, Ref.~\onlinecite{cohen-tannoudji77} for a discussion of this point).\\
\newline
\noindent\textbf{Electron in a Parabolic Quantum Dot --- Fock-Darwin States}\\
In the Symmetric gauge, the addition of an isotropic parabolic confining potential can be studied analytically. This results in so-called Fock-Darwin states.\cite{rontani99}
Fock first addressed the problem of the eigenstates of a charged particle in a uniform magnetic field in 1928.\cite{fock28} He discovered a generally non-degenerate
spectrum, particularly because the charged particle was confined to a parabolic potential. Somewhat later Landau\cite{landau30} examined a similar (simpler) problem in free space,
and found a degenerate spectrum given in Eq.~(\ref{energy_landau_level}), which now bears his name. A year later, Darwin\cite{darwin31} independently obtained results similar to those of Fock, and figures displaying energy levels as a function of applied field are now called Fock-Darwin Spectra. The presence of a confining potential in the form of a parabolic trap requires very little additional work, so we include a description of this case in this Appendix, also featured, for example, in Ref.~[\onlinecite{rontani99}].

\begin{figure}[H]
\begin{center}
\includegraphics[scale=0.55]{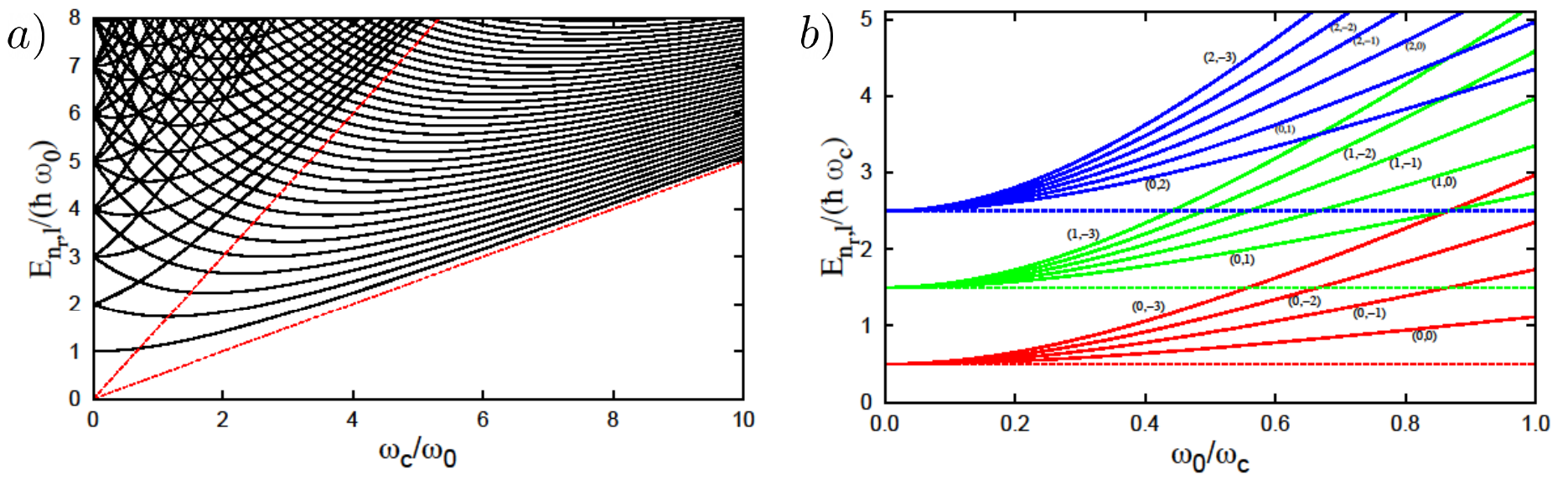}
\end{center}
\caption{ \textbf{a)} Fock-Darwin spectra obtained by using the symmetric gauge for a charged particle in a parabolic trap, i.e. Eq.~(\ref{energies_parabolic}).
A defining feature of this spectrum is the 'level crossing' phenomenon that the confined energies exhibit between free space Landau level energies, as discussed in Ref.~\onlinecite{rontani99}. The dashed (red) lines correspond to the free Landau levels. Note however, that
there is no condensation of levels as the field increases, no matter how strong the field. This is in contrast to what happens with more confining traps, to be discussed later. \textbf{b)} The same Fock-Darwin spectra obtained as in a)
but now $E_{n_r,\ell}/(\hbar \omega_c)$ is plotted vs $\omega_0/\omega_c$ for some selected levels. The numbers in brackets label the $(n_r,\ell)$ quantum numbers. The curves of a given colour all emerge from the horizontal dashed line drawn for that same colour; these denote the infinitely degenerate free Landau level energies ($\omega_0 = 0$), while the labelled curves 
of the same colour indicate how the degeneracy is broken with increasing confinement (increasing $\omega_0$). It is clear from those levels drawn here that the confinement plays a more important role for negative values of $\ell$ and increasing values of $| \ell |$. This plot also exhibits level crossing, as indicated by the intersection of curves of different color. Note that we could include all negative $\ell$ states, and these would rise even more than those shown as the confinement potential becomes stiffer. But at the far left of
Fig.~\ref{fock_darwin} b) ($\omega_0 \rightarrow 0$), we approach the free space limit, where the levels all become degenerate 
(this limit is indicated by the dashed horizontal lines).}
\label{fock_darwin}
\end{figure}

In the Symmetric gauge, with an additional confining potential of the form $\frac{1}{2}m\omega_0^2r^2/2$, the Hamiltonian given in Eq.~(\ref{ham_symm1}) becomes
\begin{equation}
H=\frac{\mbf{p}^2}{2m}+\frac{\omega_c}{2}L_z+\frac{1}{2}m\Omega^2r^2,
\label{ham_parabolic}
\end{equation}
where $\Omega \equiv \sqrt{\omega_0^2+\omega_c^2/4}$. With this parabolic confinement, the eigenenergies are
\begin{equation}
    E_{n_r,l}=2\hbar\Omega \biggl(n_r + \frac{1}{2} [ |\ell | + \frac{\omega_c}{2 \Omega} \ell] + \frac{1}{2} \biggr).
    \label{energies_parabolic}
\end{equation}
Figure~\ref{fock_darwin}(a) shows a typical Fock-Darwin plot.
A variation on this plot is shown in Fig.~\ref{fock_darwin}(b). As in Fig.~\ref{fock_darwin}(a), Eq.~(\ref{energies_parabolic}) is used for the energies, but now they are normalized to the energy $\hbar \omega_c$, and are plotted as a function of the (normalized) confining potential characteristic frequency.

\subsection{\black{The Landau Gauge}}

\noindent\textbf{Eigenstates and Eigenenergies}\\
The derivation of the eigenstates/eigenenergies in the Landau gauge is much more straightfoward than their symmetric gauge counterparts. For this reason, details of the derivation will be kept brief.
Later on we will illustrate explicitly for confined geometries that correct gauge-invariant properties are obtained for various gauge choices. Here we will not demonstrate this explicitly, but instead point
out that differences are apparent in the free space case, and these differences are clearly rooted in the infinite degeneracy that occurs in this case. We therefore proceed in this section with the Landau gauge discussed
in the Introduction, ${\vec{\bf A}}_1 = xB \hat{\bf y}$.
Now our Hamiltonian is
\begin{equation}
    H=\frac{1}{2m}\left[p_x^2+\pi_y^2\right]=\frac{1}{2m}\left[p_x^2+(p_y+eBx)^2\right]
\end{equation}
where $\pi_y$ is the kinetic momentum operator in the $\boldsymbol{\hat{y}}$ direction.
Immediately, we see that $[H,p_y]=0$, so we can separate our eigenstate $\Psi$ as
\begin{equation}
    \Psi(x,y)\propto e^{ik_yy}\psi(x),
\end{equation}
where the operator $p_y$ is replaced by its eigenvalue $\hbar k_y$. Substituting this into the Schr\"odinger Equation leads to the one dimensional differential equation
\begin{equation}
        \left[\frac{-\hbar^2}{2m}\frac{d^2}{dx^2}+\frac{1}{2}m\omega_c^2(x+k_yl_B^2)^2\right]\psi(x)=E\psi(x),
\end{equation}
which is the usual 1D harmonic oscillator (albeit not centred at zero) equation. The derivation to the solution of the 1D QHO is very well known (see, for example, Ref.~\onlinecite{griffiths18}), and will not be repeated here; it is
\begin{equation}
    \Psi_{n,k_y}(x,y)=e^{ik_yy}\frac{1}{\sqrt{2^{n}n!
    \sqrt{\pi}l_B}}e^{-\frac{(x+k_yl_B^2)^2}{2l_B^2}}H_{n}\left[\frac{x+k_yl_B^2}{l_B}\right]
    \label{psi_landau}
\end{equation}
where $H_n(z)$ are the Hermite polynomials.\footnote{Like the Laguerre polynomials, different definitions exist for the Hermite polynomials that affect normalization. In Eq.~(\ref{psi_landau}), we use the physicists' convention, where $H_{n}(z)$ has a coefficient of $z^{n}$ equal to $2^{n},$ i.e. $H_{0}(z)=1, H_{1}(z)=2 z, H_{2}(z)=4 z^{2}-2,$ etc. This definition can be found in Griffith's Introduction to Quantum Mechanics.\cite{griffiths18}}
This method of solution has introduced two quantum numbers, $k_y$, a real number, and $n$, a non-negative integer. This latter quantum number counts the number of nodes of the probability density.
The corresponding energies are given by:
\begin{equation}
    E_{n,k_y} = \hbar\omega_c\left(n+\frac{1}{2}\right).
\end{equation}
Note the lack of dependence on the quantum number $k_y$. Consequently, the energies are infinitely degenerate for every non-negative integer $n$. Here, $n$ corresponds to the Landau level quantum number $n_L$ introduced for the symmetric gauge energies in Eq.~(\ref{energy_landau_level}). Note that the wave functions are one-dimensional-like and certainly don't resemble those obtained previously. The reason this is possible is
because of the degeneracy present (energy does not depend on $k_y$), which is further discussed in the main body of the text.\\
\newline
\textbf{Lowest Landau Level}\\
In the Landau gauge, the $LLL$ wavefunction must have $n=0$ for any $k_y\in\mathbb{R}$, giving us:
\begin{equation}
    \Psi_{LLL}(x,y)=e^{ik_yy}\frac{e^{-\frac{(x+k_yl_B^2)^2}{2l_B^2}}}{\sqrt{\sqrt{\pi}l_B}}.
\end{equation}
Usually periodic boundary conditions are imposed in the $y$-direction, resulting in the requirement that $k_y = 2\pi m_y/a$, where $a$ is the length of the sample (i.e. ``ribbon'') in the $y$-direction and now $m_y$ is an integer. We can estimate the degeneracy here as well. The magnetic flux $\Phi_B$ between the centres of two successive $m_y$ eigenstates $\Psi_{n,m_y+1}$ and $\Psi_{n,m_y}$ is
\begin{equation}
   \Phi_B=Ba(x_{k+1}-x_k)=Ba\left(\frac{2\pi (m_y+1)}{a}l^2_B-\frac{2\pi m_y}{a}l^2_B\right)=B\frac{2\pi\hbar}{eB}=\frac{h}{e}=\Phi_0,
   \label{degeneracy_landau}
\end{equation}
so, as in the Symmetric gauge, each Landau gauge eigenstate contains roughly a single quantum of flux, $\Phi_0$. Thus, the expected degeneracy $G$ in a finite system of area $A$ is $G = BA/(h/e) = \Phi/\Phi_0$, as before.

\bibliography{References}

\end{document}